\documentclass[]{aa}
\usepackage[]{graphicx,amsmath}
\usepackage{txfonts}
\usepackage{color}
\newcommand{\bd}[1]{\mbox{\boldmath $#1$}}
\newcommand{\brafracket}[2]{\left( \frac{#1}{#2} \right)}
\newcommand{\sub }[1]{_{\textrm{#1}}}
\newcommand{\Msun}{\rm M_\odot}
\newcommand{\Mjup}{\rm M_{Jup}}
\newcommand{\Rsun}{\rm R_\odot}
\newcommand{\Lsun}{\rm L_\odot}
\def\wig#1{\mathrel{\hbox{\hbox to 0pt{%
          \lower.6ex\hbox{$\sim$}\hss}\raise.4ex\hbox{$#1$}}}}
\defcitealias{Hosokawa+11}{HOK11}
\defcitealias{BC10}{BC10}
\defcitealias{BCG09}{BCG09}
\defcitealias{BVC12}{BVC12}
\usepackage[ 
	colorlinks = true,
	linkcolor = blue,
        urlcolor  = blue,
        citecolor = blue,
        anchorcolor = blue
]{hyperref}
\bibpunct{(}{)}{;}{a}{}{,} % to follow the A&A style

\begin{document}

   \title{Revisiting the pre-main-sequence evolution of stars}

   \subtitle{I. Importance of accretion efficiency and deuterium abundance
   \thanks{
The isochrones are only available at the CDS via anonymous ftp to cdsarc.u-strasbg.fr (130.79.128.5) or via \ \ \ \ \ \ \ \ \ \ \ \ \ \ \ \ \ \ \ \ \ \ \ \ \ \ \ \ \ \ \ \ \ \
\href{http://cdsarc.u-strasbg.fr/viz-bin/qcat?J/A+A/599/A49}{http://cdsarc.u-strasbg.fr/viz-bin/qcat?J/A+A/599/A49}
   }
   }

   \author{Masanobu Kunitomo\inst{\ref{inst1}},
          Tristan Guillot\inst{\ref{inst2}},
          Taku Takeuchi,\inst{\ref{inst3}}
          \fnmsep\thanks{Present affiliation: Sanoh Industrial Co., Ltd., Japan} 
         \and
         Shigeru Ida\inst{\ref{inst4}}
          }

   \institute{Department of Physics, 
                        Nagoya University, Furo-cho, Chikusa-ku, Nagoya, Aichi 464-8602, Japan\label{inst1}\\
                        \email{kunitomo@nagoya-u.jp}
                \and
                     Universit\'e de Nice-Sophia Antipolis,
             Observatoire de la C\^ote d'Azur,
             CNRS UMR 7293,
             06304 Nice CEDEX 04, France\label{inst2}
                \and
                         Department of Earth and Planetary Sciences, 
                         Tokyo Institute of Technology,
                         2-12-1 Ookayama, Meguro-ku, Tokyo 152-8551, Japan\label{inst3}
         \and
                         Earth-Life Science Institute,
                         Tokyo Institute of Technology,
                         2-12-1 Ookayama, Meguro-ku, Tokyo 152-8551, Japan\label{inst4}
}
   \date{Received 5 February 2016 / Accepted 6 December 2016}

% \abstract{}{}{}{}{} 
% 5 {} token are mandatory
 
  \abstract
   {
   {Protostars grow from the first formation of a small seed and subsequent accretion of material.}
   Recent theoretical work has shown that the pre-main-sequence (PMS) evolution of stars is much more complex than previously envisioned.
        {Instead of the traditional {steady, one-dimensional} solution,}       
   accretion may be episodic and not necessarily symmetrical, thereby affecting the energy deposited inside the star and its interior structure.
   }
   {Given this new framework, we want to understand what controls the evolution of accreting stars.
   }
   {We use the MESA stellar evolution code with various sets of conditions. In particular, we {{account for the (unknown)}} efficiency of accretion in burying gravitational energy into the protostar {{through a parameter, $\xi$, and we vary}} the amount of deuterium present.}
   {We confirm the findings of previous works that, in terms of evolutionary tracks on the Hertzsprung-Russell (H-R) diagram, the evolution changes significantly with the amount of energy that is lost during accretion. 
   We find that deuterium burning also regulates the PMS evolution. 
   In the low-entropy accretion {{scenario}}, the evolutionary tracks in the H-R diagram are significantly different from the classical tracks and are sensitive to the deuterium content. A comparison of theoretical evolutionary tracks and observations allows us to exclude some cold accretion models {{($\xi\sim 0$)}} with low deuterium abundances.
   }
   {{{We confirm that the luminosity spread seen in clusters can be explained by models with a somewhat inefficient injection of accretion heat. The resulting evolutionary tracks then become sensitive to the accretion heat efficiency, initial core entropy, and deuterium content. In this context, we predict that clusters with a higher D/H ratio should have less scatter in luminosity than clusters with a smaller D/H. }}
        {{Future work on this issue should include}} radiation-hydrodynamic simulations to determine the efficiency of accretion heating and further observations to investigate the deuterium content in star-forming regions.
   }


   \maketitle
%
%________________________________________________________________

\section{Introduction} \label{sec:intro}
Since the pioneering work of {{\cite{Hayashi61},}} the first phase of stellar evolution, the so-called pre-main-sequence or PMS, is generally considered with the following simple approach: 
{{A star is formed with a large radius and contracts isotropically,}}
and a huge release of gravitational potential energy heats the interior and yields a similarly large luminosity, which ensures convection in most of the interior of the star. As the star heats up, thermonuclear reactions set in, contraction is slowed, and a radiative zone begins to grow. In standard models for the contraction of a solar mass star, the growth of the radiative zone starts after 2 million years and the star reaches the main sequence (with an outer convective zone that is only about 2.5\% of the total mass of the present Sun) after about 40 million years. This long phase, in which the star possesses a very deep convective zone or is even fully convective, almost guarantees a homogeneous composition in the stellar interior owing to extremely fast mixing in convective zones. 
A large portion of the gravitational energy of the accreted material is supposed to be given to the star, ensuring a large radius and important luminosity {{\citep[e.g.,][]{SST80I}.}}

Contrary to that ideal picture, a number of studies have revealed that the PMS evolution can be strongly affected by the way material is accreted onto the star {{during the accretion phase}}. If material loses entropy before or during the accretion onto the star, it can grow from a small radius and avoid the large quasi-static contraction phase \citep[e.g.,][]{Mercer-Smith+84,Palla+Stahler92,Hartmann+97,BCG09,Hosokawa+11,Dunham+Vorobyov12}. This has strong consequences for the stellar evolutionary tracks and therefore the inferred physical properties of the star \citep{BCG09, Hosokawa+11, BVC12}. Finally, it controls the level to which planet formation affects {{stellar surface composition}}, as suggested from observations \citep{Ramirez+09,Ramirez+11,Melendez+09} or theoretical models \citep{Chambers10,Guillot+14}.

In this article, we wish to understand what controls the evolutionary tracks on the PMS, explain apparent differences seen in published results, and obtain limits on physically plausible evolutionary tracks from observational constraints. In a subsequent article, we will attempt to understand what controls the evolution of radiative and convective zones in PMS stars.

This paper is organized as follows. In Sect.~\ref{sec:method}, we describe our physical model and computation method for simulating the PMS evolution including accretion. 
In Sect.~\ref{sec:cold}, we examine how, and to what extent, energy losses during accretion control PMS evolution. We also identify deuterium as playing a leading role in this evolution phase despite its fast burning nature.
In Sect.~\ref{sec:hot}, we explore the dependence on the entropy of accreting matter.
In Sect.~\ref{sec:HR}, we compile the results of Sects.~\ref{sec:cold} and \ref{sec:hot} to evaluate observational constraints and observational consequences. Our results are summarized in Sect.~\ref{sec:conclusion}.

\section{Method}\label{sec:method}

%________________________________________________________________
\subsection{Stellar evolution code}\label{sec:code}
Our evolution models are calculated for spherically symmetric single stars without strong magnetic fields or rotation, but including accretion from a small initial seed.
We use the one-dimensional stellar evolution code MESA version 6596 \citep[][]{Paxton+11,Paxton+13,Paxton+15} and refer to the Paxton et al. papers for full details of the computational method.

The code numerically solves the equations of continuity, momentum, energy, temperature gradient, and composition.
We assume the hydrostatic equilibrium for the momentum equation. 
The temperature gradient is determined by the mixing length theory of \citet{Cox+Giuli68} and \citet{Henyey+65}. The Ledoux criterion of convection is used.
The composition is changed by thermonuclear reaction and diffusion. 
The diffusion coefficient is given by the mixing length theory and overshooting of \citet{Herwig00}, in which the diffusion coefficient exponentially approaches zero in the radiative zone from the boundary with the convective zone.
We use the mixing length parameter, $\alpha\sub{MLT}=1.90506,$ which is the ratio of the mixing length to the pressure scale height, and the overshooting parameter, $f\sub{ov}=0.0119197$, which determines the extent of the overshooting region \citep[see Eq.~9 of][]{Paxton+13} (see Appendix~\ref{app:param}).
The energy equation is described in Sect.~\ref{sec:Sacc} in detail.

We use the radiative opacity of \citet[][{OP}]{Seaton05} and \citet[][]{Ferguson+05} for the temperature range below $10^{4.5}~{\rm{K}}$, and the electron conduction opacity of \citet{Cassisi+07}.
We adopt the MESA\ default version of the equation of state; we basically follow the OPAL EOS tables \citep{Rogers+Nayfonov02} and SCVH tables \citep{Saumon+95}.

%________________________________________________________________
\subsection{Thermonuclear reactions} \label{sec:nuc}
Pre-main-sequence stars ignite light elements such as deuterium and lithium when the central temperature exceeds $\sim 10^6$ and $\sim3\times10^6~\rm{K}$, respectively. 
The deuterium burning is as follows: {{$\textrm{D} + \,^1\textrm{H} \rightarrow \,^3\textrm{He} + \gamma$}}.
This is a strongly exothermic reaction, which produces $5.494~\textrm{MeV}$ via one reaction.
Although the lithium burning produces more energy by a factor of about three, it does not affect the evolution owing to a $\sim 10^3$ times smaller abundance \citep{GS98}.
We use the thermonuclear reaction rates of \citet[][{NACRE}]{Angulo+99}.
Main-sequence (MS) stars ignite hydrogen when temperature exceeds $\sim~10^7~\rm{K}$.

%________________________________________________________________
\subsection{Chemical composition} \label{sec:chem}

The initial mass fractions of hydrogen, helium, and metals are {{denoted by}} $X\sub{ini}, Y\sub{ini}$, and $Z\sub{ini}$, respectively.
We choose the input parameters that reproduce  values of the present-day Sun estimated by helioseismic and spectroscopic observations (see Appendix~\ref{app:param}). 
We conducted a $\chi^2$ test and used the following converged input parameters: $X_{\mathrm{ini}}=0.70046$, $Y_{\mathrm{ini}}=0.27948$ and $Z_{\mathrm{ini}}=0.02006$.
We assume $^3\mathrm{He}/^4\mathrm{He} = 10^{-4}$, similar to the value in the Jovian atmosphere \citep{Mahaffy+98}.
We use the composition of metals of \citet{GS98}.

As described in the previous section, the deuterium content is a key parameter for the PMS evolution. In this study we use the mass fraction of deuterium, $X\sub{D}$, and our fiducial $X\sub{D}$ is 20~ppm (parts per million, $10^{-6}$), which is an interstellar value. 
However, the interstellar $X\sub{D}$ remains uncertain. Thus we explore the consequences of varying the deuterium content in Sect.~\ref{sec:coldD}.

Finally, the accreting gas is assumed to have constant composition with time.
Importantly, this includes fresh deuterium, which is added to the outer layers.
Since deuterium is quickly burned in the stellar interior, this is an important factor governing the evolution on the PMS. The deuterium abundance profile is calculated by accounting for diffusion.

%________________________________________________________________
\subsection{Initial conditions}\label{sec:method-init}

Stars are formed by the collapse of molecular cloud cores through a number of stages \citep[e.g.,][]{Larson69,Winkler+Newman80,Inutsuka12}. As the collapse of a molecular cloud core proceeds, the central density increases and the central region becomes adiabatic. After the first hydrostatic core is formed, the dissociation of hydrogen molecules causes the second collapse after the central temperature exceeds $\sim2000~\rm{K}$. As a result of the second collapse, the central temperature increases and eventually a second hydrostatic core is formed.

In this paper, we focus on the evolution after the formation of the second core.
This second core is only about a Jovian mass ($\Mjup$) initially, and
most of the mass is thus accreted in a subsequent phase. As accretion is progressively suppressed, the star enters its quasi-contraction phase, increases its central temperature, and eventually becomes a main-sequence star. 

In this paper, we set our fiducial values of the initial mass and radius as $0.01~\rm{M_\odot}$ and $1.5~\rm{\Rsun}$, respectively, following \citet{SST80I} and \citet[][hereafter \citetalias{Hosokawa+11}]{Hosokawa+11}. This choice is essentially driven by convergence issues with MESA at low masses.
 However, our fiducial initial mass is higher than the second core mass, which is derived by radiation-hydrodynamic simulations \citep[$\sim1$--$4\,\Mjup$; e.g.,][]{Masunaga+Inutsuka00,Vaytet+13}.
{{Therefore initial seeds in the present work correspond to slightly evolved protostars from second cores.}}
As discussed by \citet[][hereafter \citetalias{BVC12}]{BVC12}, the values of the initial seed mass {{and radius set}} the entropy of the forming star and therefore its subsequent evolution. We show in Appendix~\ref{app:Mini} that for ranges of masses between 1 and $4\,\Mjup$, the range of radii when the mass attains $0.01~\rm{M_\odot}$ is between $0.25~\rm{\Rsun}$ and $1.5~\rm{\Rsun}$. Our fiducial value of $1.5~\rm{\Rsun}$ therefore corresponds to a high initial value of the entropy of the star. In Sect.~\ref{sec:cold-ini} we explore the consequences of variations of the initial radius down to $0.25~\rm{\Rsun}$.

%________________________________________________________________
\subsection{Mass accretion} \label{sec:method-Mdot}

The collapse of the molecular cloud core yields an accretion rate $\dot{M} \sim c\sub{s}^3/G$, where $c\sub{s}$ is the characteristic speed of sound in the cloud \citep[e.g.,][]{Shu77,SST80I}. For molecular cloud temperatures $T\sim 30$\,K, this implies $\dot{M}\approx 10^{-5}\,\rm M_\odot/\textrm{yr}$, which is our fiducial accretion rate.
The accretion onto the star is however mostly determined by the angular momentum of the collapsing gas and formation of a circumstellar disk \citep[e.g.,][]{Hueso+Guillot05,Vorobyov+Basu10,Inutsuka+10,Martin+12}. 
It can be strongly variable (episodic), as observed in FU Ori.

{Our fiducial accretion rate is $\dot{M}\approx 10^{-5}\,\rm M_\odot/\textrm{yr}$ but} we explore the effect of varying this rate in Sect.~\ref{sec:Mdot}.
We stop mass accretion abruptly when the stellar mass reaches the final mass, $M\sub{fin}$.

%________________________________________________________________
\subsection{Modeling the consequences of accretion}\label{sec:Sacc}

The consequence of accretion onto the forming star is of course an increase of its mass, but this accretion also modifies the stellar environment and its radiation to space and delivers energy to the star.
{{Accretion}} also delivers fresh deuterium,  which is {{an important}} source of combustible material. 

It is difficult at this point to model the problem in its full complexity, in particular, because this should involve accounting for the angular momentum of material in the molecular cloud core, its magnetic field, the presence of outflows, and the particular geometry that arises from the birth of a circumstellar disk.
Instead, we adopt a simple parametric approach inspired by the work of \citet[][hereafter \citetalias{BCG09}]{BCG09}. 

First, we assume that the specific entropy of the accreting material is the same as that at the stellar surface. 
This follows from \cite{Hartmann+97} {estimate} that, in the case of accretion from disk to protostar, any entropy (equivalent temperature) excess
would be quickly radiated to space because, especially in the presence of
an inner cavity, the optical thickness toward the stellar interior is much larger than toward the disk. 
In practice, this treatment is favorable to the computational convergence.
We point out that validating this hypothesis would require a multidimensional radiation-hydrodynamic simulation beyond the scope of the present work. 

Second, we parametrize the heat injected by the accreting material as
\begin{equation} \label{eq:Ladd}
L\sub{add}=\xi GM_\star\dot{M}/R_\star,
\end{equation}
where $M_\star$ and $R_\star$ are stellar mass and radius, respectively, and energy constraints impose that $\xi$ is such that $0\le \xi\le 1$
\footnote{
Our $\xi$ is equivalent to $\alpha \epsilon$ in \citet{BCG09,BVC12}.
}.
In situations in which radiation to space is favored (e.g., when accreting from a disk with an inner cavity), we expect a low value of $\xi$. Even when this is not the case, \citet{SST80I} show that radiative transfer considerations impose that $\xi\le 3/4$ in the spherical accretion. 
If materials accrete onto the star through an active Keplerian disk, the upper limit of $\xi$ should be 0.5 owing to the radiative cooling from the disk surface.
If the star rotates rapidly, a large fraction of gravitational energy can be stored in the rotational energy and then $\xi$ is decreased.
Therefore, in practice we opt for $\xi=0.5$ as a realistic upper limit for what we call ``{hot accretion}''.
We label simulations with $\xi=0 $ as ``{cold accretion}'' simulations and we label as ``{warm accretion}'' simulations all cases where $0<\xi<0.5$.

Third, we use two simple models to parametrize how the energy enters the stellar interior. The first is the uniform model, which is adopted by \citet[][hereafter \citetalias{BC10}]{BC10}, and in which $L\sub{add}$ is distributed uniformly and instantaneously within the entire star. The energy deposited per unit mass is thus 
\begin{equation}\label{eq:eps_uniform}
\varepsilon\sub{add}^{\rm (uniform)}=\frac{L\sub{add}}{M_\star}.
\end{equation}
The second  is the linear model, in which we consider that the accretion energy is deposited only in an outer layer of relative mass $m\sub{ke}$, such that {the accretion energy is zero in the layer from the stellar center to relative mass $(1-m\sub{ke})$} and that it increases linearly with mass until the photosphere.
With this assumption, the energy deposited per unit mass is
\begin{equation}\label{eq:eps_linear}
\varepsilon\sub{add}^{\rm (linear)}=\frac{L\sub{add}}{M_\star}{\rm Max}\left[0,\frac{2}{m\sub{ke}^2}\left( \frac{M_{r}}{M_\star}-1+m\sub{ke} \right) \right].
\end{equation}
where $M_{r}$ is the mass coordinate and $m\sub{ke}$ is expressed as a fraction of the total mass of the star at time $t$ so that $0< m\sub{ke}\le 1$. 
 \
In both cases, $\varepsilon\sub{add}$ satisfy the relation $\int \varepsilon\sub{add} dM=L\sub{add}$. The energy conservation equation thus is written 
\begin{equation}\label{eq:dLdM}
  \frac{\partial L}{\partial M_{r}} = \varepsilon_{\mathrm{nuc}}-\left( T\frac{\partial S}{\partial t} \right)_{M_r}+\varepsilon_{\mathrm{add}},
\end{equation}
where $L$ is the luminosity, $T$ its temperature, $S$ its specific entropy, $t$ time, and $\varepsilon\sub{nuc}$ is the energy production rate from thermonuclear reactions in the shell.

The uniform model thus corresponds to the case studied by \citetalias{BC10} in which $L\sub{add}$ is distributed uniformly and instantaneously within the entire star. In the linear model, we consider that the accretion energy is deposited preferentially in the outer layers of the star and that this energy deposition is linear in mass over a shell of mass $m\sub{ke}$ (with mass expressed as a fraction of the total mass of the star at time $t$).

BC10 justify the uniform model by the fact that low-mass ($\lesssim 2~\Msun$) stars are fully convective during their PMS stage and that convection may rapidly deliver the accretion energy into the deep interior. We believe that its efficient transport of energy (or equivalently entropy) is probably unrealistic owing to the difficulty in transporting energy into a star. Furthermore, the generation of heat at the surface of accretion flows can potentially hinder convection, thereby preventing any uniform entropy mixing, as recently found by two-dimensional hydrodynamic simulations with  high accretion rates \citep{Geroux+16}. The linear model is hence a highly simplified but useful parametrization that allows us to explore the consequences of a penetration of the accretion energy only to a fraction of the radius of a star.

In the case of hot accretion, \citetalias{Hosokawa+11} adopt a different approach; they do not consider that accretion energy can be transported inward but instead adopt an outer boundary condition to account for the heat generated by accretion \citep[see][]{Hosokawa+Omukai09}. In principle, this should be a better approach. However, the approach remains one-dimensional and is not parametrized. It does not include the possibility that part of the energy may be transported to deeper levels in an non-radiative way. We therefore believe that the different approaches are complementary. 

Numerically, evaluating the thermodynamic state of the accreting material requires careful consideration. 
The mass increase is performed within MESA using an ``Eulerian'' scheme \citep{Sugimoto+Nomoto75}.
MESA originally treated {the thermodynamic state of the accreting material} with compressional heating \citep{Townsley+Bildsten04,Paxton+13}. However, this is not suitable for rapid accretion whose timescale is shorter than the thermal relaxation timescale. Thus, the new scheme was implemented in version 5527 and is used in this study.
The detailed treatment is described in Sect.~7 of \citet{Paxton+15}.

%________________________________________________________________
\subsection{Outer boundary condition} \label{sec:BC}
The pressure $P\sub{s}$ and temperature $T\sub{s}$ at the outer boundary is specified by the interpolation of a model atmosphere which is calculated with the assumption that materials accrete onto a small fraction of the stellar surface from the thin disk and do not affect the properties of the photosphere. 

We found that the choice of the model atmosphere table directly affects the convergence of the calculations. We hence selected the following two tables depending on stellar mass and radius: \\
For stars such that $M_\star \geq 1~\Msun$ and $R_\star > 0.7~\Rsun$, we adopted a ``photospheric'' table for an optical depth $\tau\sub{s}=2/3$ and $T\sub{s}=T\sub{eff}$ and a surface pressure $P\sub{s}$ set by the model atmospheres PHOENIX \citep[][]{Hauschildt+99a, Hauschildt+99b} and ATLAS9 \citep[][]{Castelli+Kurucz04}. For less massive or smaller stars, we used  the ``$\tau\sub{s}=100$'' model, which specifies $P\sub{s}$ and $T\sub{s}$ at $\tau\sub{s}=100$ from the ATLAS9 and COND \citep[][]{Allard+01} model atmospheres.
The impact of the switch of the boundary condition on the stellar structure and evolution is negligible.

%________________________________________________________________
\subsection{Adopted parameters and comparison with previous studies} \label{sec:input_param}

Table~\ref{Table:param} summarizes the values of our main parameters. Our fiducial values are shown in boldface. We also provide the range of values that we consider in Sects.~\ref{sec:cold}--\ref{sec:HR} when we seek to explore the consequences of deviations of some parameters from the fiducial values. In particular, we explore the dependence on the deuterium mass mixing ratio, something which had not been considered so far.

%__________________________________________________ One column table
   \begin{table*}[!htb]
        \begin{center} \centering
      \caption[]{Input parameters.}
         \label{Table:param}
         \small
         \begin{tabular}{p{0.3\linewidth}lll}
            \hline
            \hline
            Quantity      &  \mbox{This study} &   \mbox{\citetalias{BC10}}\tablefootmark{a} &  \mbox{\citetalias{Hosokawa+11}}\tablefootmark{b} \\
            \noalign{\smallskip}
            \hline
        Initial mass, $M\sub{ini}~[\Msun]$ & $\bd{0.01}$ & $0.0095$ &  $0.01$ \\
                Initial radius, $R\sub{ini}~[\Rsun]$ &  $\bd{1.5}$ & N/A        &       $1.5$ \\
                                        & $(0.25$--$3)$  & &  \\
                Final mass, $M\sub{fin}~[\Msun]$ & $\bd{1}$      & $1$ & $0.9$\\
                                        & $(0.025$--$1.5)$    &  & \\
                Mass accretion rate, $\dot{M}~[\Msun/\rm{yr}]$ &  $\bd{10^{-5}}\tablefootmark{c}$ & \textrm{Episodic} & $10^{-5}$ \\
                                        & $(10^{-6}$--$10^{-4}$ or episodic)\tablefootmark{d}  &  & \\
                Deuterium mass fraction, $X\sub{D}~[\textrm{ppm}]$ &  \bd{20} & $20$ & $35$\\
                                        & $(0$--$40)$    &  & \\
                Heat injection efficiency, $\xi$ & \bd{0} \textbf{(Cold accretion)}   & $0$ & $0$\\
                                        & $(0$--$0.5)$   &  & \\
                Heat injection region & \textbf{Uniform}   & $-$ & $-$\\
                                        & $(m\sub{ke}=0.1, 1)$ &  &   \\
                Mixing length parameter, $\alpha\sub{MLT}$ & \bd{1.905}  & $1.0$ & $1.5$\\
                    \noalign{\smallskip}
            \hline
                    \noalign{\smallskip}
         \end{tabular} 
         \end{center}   
                    {\textbf{Notes.}              The values used by default in this work are highlighted in boldface. The other values described below are used in Sects.~\ref{sec:ini} ($R\sub{ini}$), \ref{sec:Mfin} ($M\sub{fin}$), \ref{sec:Mdot}--\ref{sec:episodic} ($\dot{M}$), \ref{sec:coldD} ($X\sub{D}$), \ref{sec:xi} ($\xi$), and \ref{sec:mke} ($m\sub{ke}$).  
                $^{(a)}$ Parameter values corresponding to the blue long-dashed line in Fig.~2 of \citet{BC10}. $^{(b)}$ Parameter values for the ``mC5-C'' case of \citet{Hosokawa+11}. $^{(c)}$ The duration of accreting phase is $0.99 \times 10^{5}~{\textrm{yr}}$ for the fiducial values of the accretion rate, initial mass, and final mass. $^{(d)}$ See Sect.~\ref{sec:episodic}.
}
\normalsize
   \end{table*}

We compare our results with two particular evolutionary models from previous studies (\citetalias{BC10} and \citetalias{Hosokawa+11}) {{in Sect.~\ref{sec:comp_prev}}}.

%________________________________________________________________
\section{Evolution in the context of a ``cold'' accretion} \label{sec:cold}
\label{sec:typical}

In this section we describe the resulting PMS evolution in the context of a ``cold'' accretion of material, i.e., in the extreme case when all of the accretion energy is radiated away (i.e., $\xi=0$ in Eq.~\ref{eq:Ladd}).

\begin{figure}[!tb]
  \begin{center}
    \includegraphics[width=\hsize,keepaspectratio,clip]{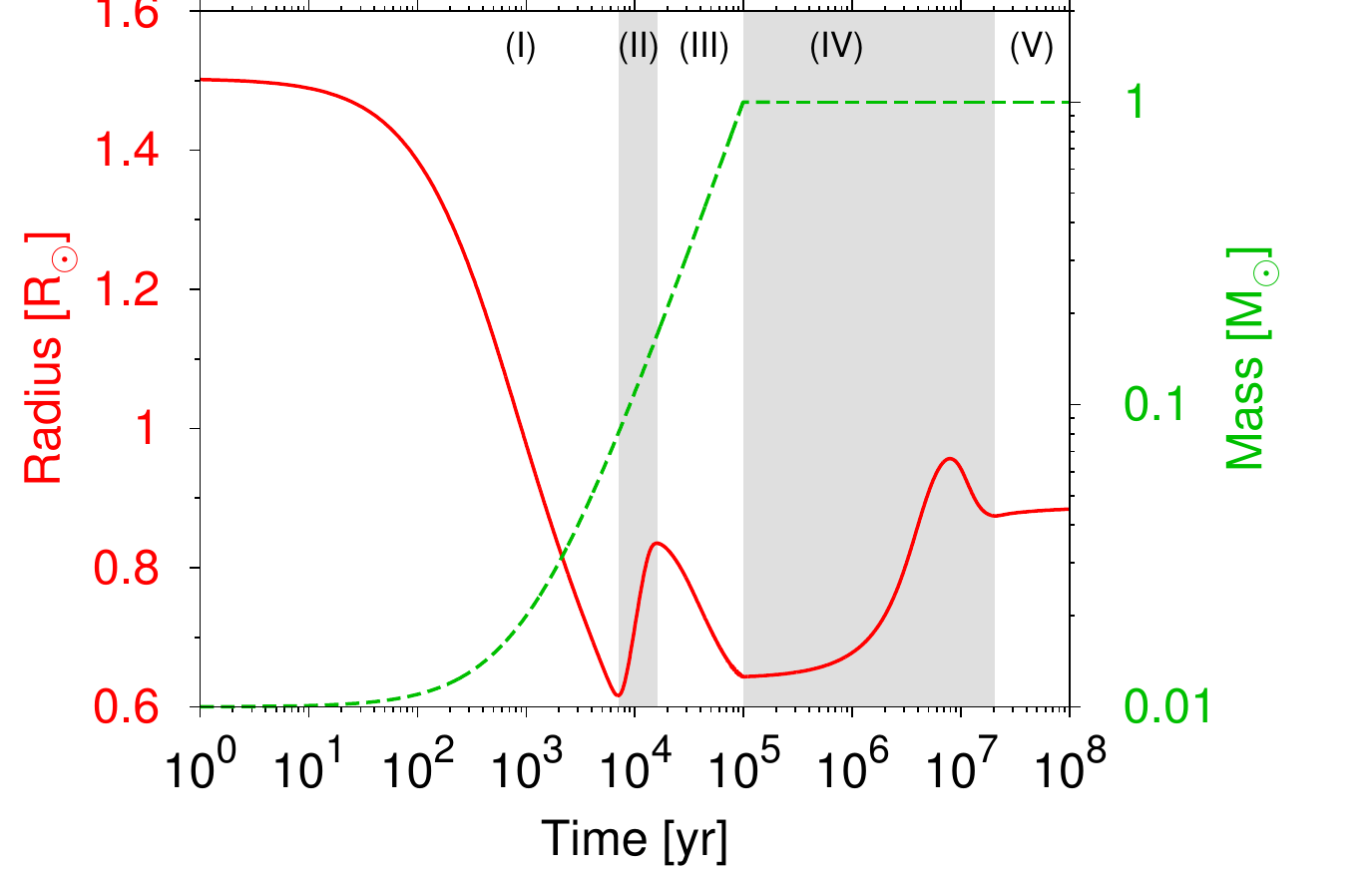}
        \caption{\small{Evolution of radius (solid line) and mass (dashed) 
        {in the cold accretion limit ($\xi=0$) and under the fiducial conditions listed in} Table~\ref{Table:param}. 
        {{The five main evolution phases are labeled as follows}}: (I) the contraction phase, (II) deuterium-burning phase, (III) second contraction phase until the accretion is completed, (IV) swelling phase, and (V) main sequence from $\sim 2\times10^7~{\textrm{yr}}$.
        }}\label{fig:1d-5t-R}
    \end{center}
\end{figure}

{{
Figure~\ref{fig:1d-5t-R} shows {how the radius changes with time in the cold accretion case and with our fiducial settings (see Table~\ref{Table:param}).
As described in more detail in Appendix~\ref{sec:overview}, the evolution} can be split into five phases: (I) the contraction phase, (II) the deuterium-burning phase, (III) the second contraction phase, (IV) the swelling phase, and (V) the main sequence.
{The contraction seen in phase (I) is induced by mass accretion with the same entropy as the star. Phase (II) begins when the internal temperature becomes high enough to start burning pre-existent deuterium, resulting in an expansion. With a decrease in the deuterium content, we enter phase (III) in which mass accretion again dominates over deuterium burning and the star contracts. After mass accretion is completed, the star expands owing to a change of its internal entropy profile. Eventually, after $\sim 2 \times 10^7~{\textrm{yr}}$, it enters the main sequence. Starting from a different initial condition (e.g., a smaller initial radius) does not change the sequence, but it alters the duration and characteristics of the star before it reaches the main sequence.}
}}

{{
After {presenting the main differences between classical evolution models and those in the context of cold accretion} (Sect.\,\ref{sec:comp_classical}), we {examine} the influence of mass accretion rate, deuterium content, final stellar mass, and initial conditions in {Sects.\,\ref{sec:Mdot}--\ref{sec:Mfin}}.
{We compare our results against previous studies} in Sect.~\ref{sec:comp_prev}.
}}

%%%%%%
%%%%%%

%_________________________________ 
\subsection{{Cold versus classical models}\label{sec:comp_classical}}
\begin{figure}[!tb]
  \begin{center}
    \includegraphics[width=\hsize,keepaspectratio,clip]{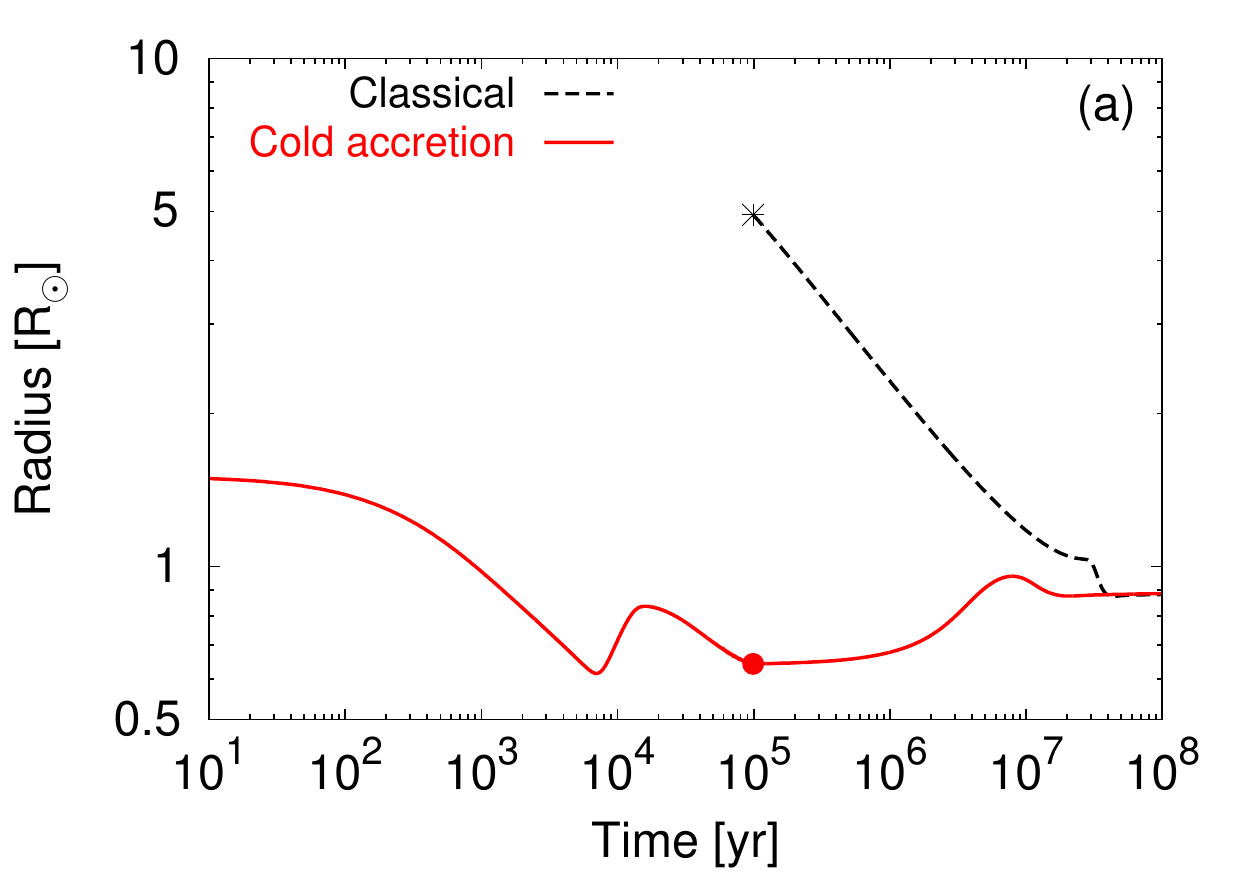}
    \includegraphics[width=\hsize,keepaspectratio,clip]{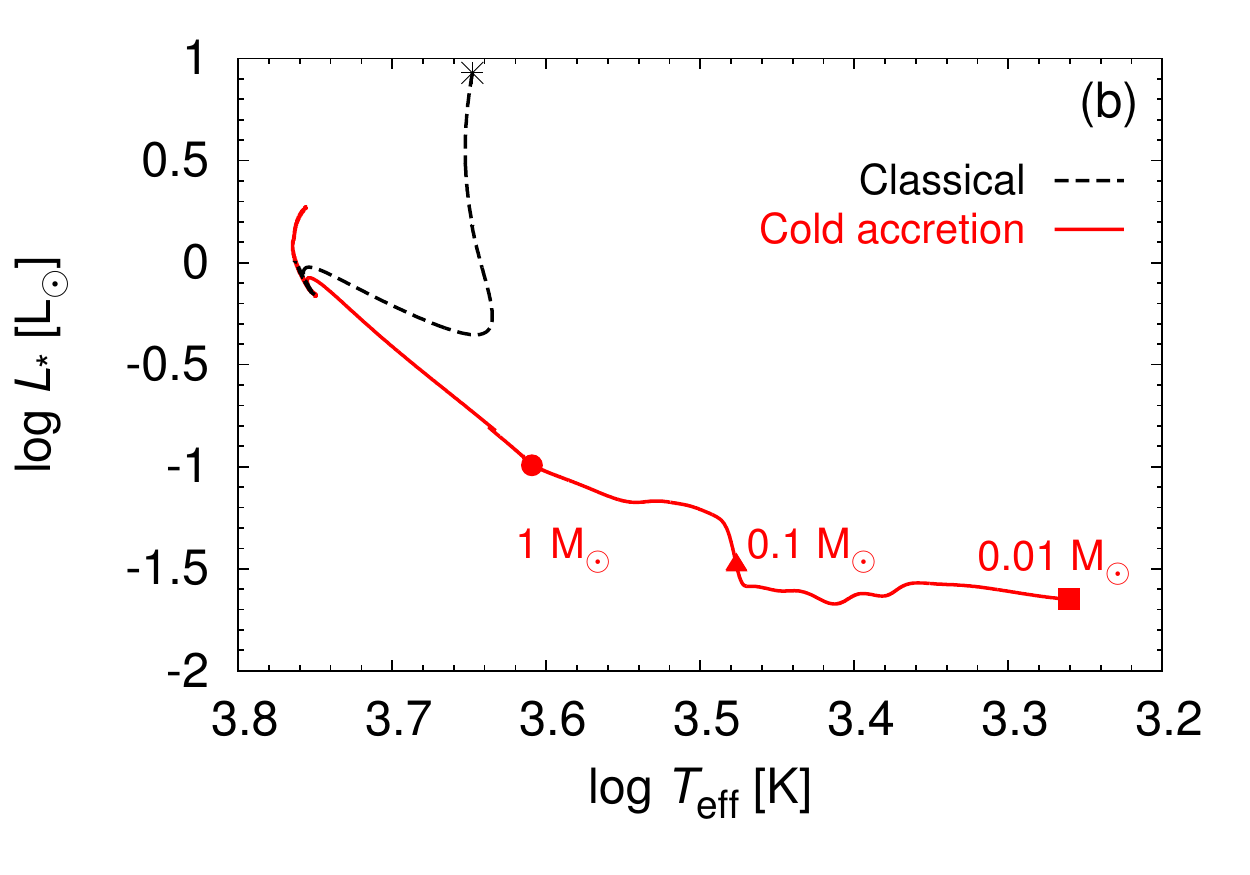}
        \caption{\small{
        Radius evolution (\textit{top panel}) and evolutionary tracks in the H-R diagram (\textit{bottom}) comparing the cold accretion (solid line) with the classical $1~\Msun$ evolution (dashed).
        The filled square indicates the initial location,  the triangle indicates the stellar mass at 0.1~$\Msun$, and the circle indicates the end of accretion.
The non-accreting classical evolution, whose initial radius is assumed to be $4.92~\Rsun$ as shown by the asterisks, starts at $0.99\times10^5~{\textrm{yr}}$ to compare the evolution.
        At $10^5~{\textrm{yr}}$ years, the radii are about one order of magnitude different.
        In the classical model the star evolves along the Hayashi and Henyey tracks, whereas in the cold accretion model the evolution is roughly horizontal in the H-R diagram.
        Although the deuterium burning slightly increases the luminosity where $\log T\sub{eff} \simeq 3.5$, it is much smaller than the Henyey track of the non-accreting $1~\Msun$ star.
                }}\label{fig:comp}
    \end{center}
\end{figure}

We now compare the radius evolution in the case of cold steady accretion with that in the classical model.
Figure~\ref{fig:comp}a shows that the stellar radius in the case of low-entropy accretion is one order of magnitude smaller than in the classical evolution case. Classically, a spherically symmetrical accretion prevents radiative losses implying a large stellar entropy and therefore a large radius. 
{{Following \cite{Stahler+Palla05}, we adopt for the classical evolution case an initial radius of $4.92\,\Rsun$ at $10^5\,{\textrm{yr}}$. 
}}
When radiative cooling of accreting materials is allowed, the resulting specific entropy is much smaller ,which yields a smaller protostar. 
Moreover, in the case of the low-entropy accretion, the MS starts at about 20 million years, which is faster than the classical evolution by a factor of about two.
This difference comes from the smaller thermal timescale in the stellar interior because of the smaller radius in the case of low-entropy accretion.

%%%%%%
%%%%%%

The pre-main-sequence evolutionary tracks for a 1\,M$_\odot$ star in the classical, non-accreting case and in the cold accretion scenario are shown in Fig.~\ref{fig:comp}b.
The tracks are drastically different. In the classical case, the star starts with a high entropy, then evolves {along} the almost vertical \textit{Hayashi track} until it reaches the horizontal \textit{Henyey track} in the Hertzsprung-Russell (H-R) diagram. In the cold accretion scenario, the initial evolution in the H-R diagram is nearly horizontal because the accretion does not inject energy and both the entropy and intrinsic luminosity remain small. The implications of these results, including the dependence on $\xi$, are discussed in Sect.~\ref{sec:HR}.

We point out that the evolutionary track in the cold accretion case in Fig.~\ref{fig:comp}b does not include the accretion luminosity. 
Once the star moves out of the class I phase and becomes a visible T-Tauri star, its accretion luminosity is easily distinguished from its intrinsic luminosity in the spectral energy distribution because it is emitted in the UV or X-rays rather than in the visible. However, in the embedded phase, the accretion luminosity is reabsorbed by the surrounding molecular cloud and re-emitted at longer wavelengths, making the distinction with the intrinsic luminosity impossible. In that case, its location in the H-R diagram should be shifted upward \citepalias{Hosokawa+11}.

{{
In the following three sections, we explore the consequences of changing the accretion history,  deuterium content, initial entropy, and final mass. The other parameters are set to be the fiducial values listed in Table~\ref{Table:param}.
}}

%_________________________________ 
\subsection{Dependence on mass accretion rate}\label{sec:Mdot}\label{sec:episodic}

As described in Sect.~\ref{sec:method-Mdot}, the accretion rate onto the star can be highly time dependent and variable from one star to the next. 
We calculate the evolution varying mass accretion rate,
ranging from $10^{-4}$ to $10^{-6}~\Msun/{\textrm{yr,}}$
and time variability; i.e., the episodic accretion following the pioneering works of \citetalias{BCG09}, \citetalias{BC10} and \citetalias{Hosokawa+11}.
For episodic accretion, we adopt the parameters of \citetalias{BC10} on the basis of hydrodynamic simulations by \citet{VB05}.

We confirm the findings of \citetalias{BCG09} and \citetalias{Hosokawa+11} that the variation of the accretion rate only has a moderate impact on the evolution. The difference in radii for the same mass is at most $0.2~\Rsun$ even for accretion rates that differ by two orders of magnitude. 
As described afterward, this is much smaller than the difference from other effects (the deuterium content $X\sub{D}$ and the heat injection efficiency $\xi$). 
The evolutionary tracks are also hardly affected by the variation of the accretion history.
We stress that rather than accretion history, it is the entropy of the accreted material that matters.
We therefore choose to fix the mass accretion rate to $10^{-5}~\Msun/\rm{yr}$ from now on.

%_______________________________________________
\subsection{Dependence on deuterium content}\label{sec:coldD}

The PMS evolution is controlled by deuterium burning when it occurs vigorously (i.e., during phase II in Sect.~\ref{sec:typical}).
The deuterium content can largely differ from star to star because of galactic evolution or the local environment. We now explore how PMS evolution is affected by deuterium content.

This was already investigated by \citet{Stahler88} who concluded that the radius is only moderately affected by deuterium content.
For example, when stars become $1~\Msun$ the radii are only increased by a factor of $\simeq1.6$ from a deuterium-free star to the case with $X\sub{D}=35~\mathrm{ppm}$. However, since they assumed a spherical accretion, this study is to be re-examined in the framework of low-entropy accretion.

First we summarize the currently available values of the deuterium content\footnote{These values are estimated using the number ratio of the deuterium over hydrogen, (D/H). Since the hydrogen mass fraction $X\simeq 0.7$ and the mass ratio of deuterium to hydrogen is about two, $X\sub{D}\simeq1.4 {\mathrm{(D/H)}}$.}. 
The primordial value at the beginning of the universe is $X\sub{D,prim}=35.8\pm2.5$~ppm \citep{Steigman06}, 
the indirectly\footnote{
Two methods may be used to constrain the protosolar deuterium abundance: using either the deuterium content in the Jovian atmosphere \citep{Lellouch+01,Guillot99} or the enhanced $^3$He measured in the solar wind and compared to the Jovian atmosphere \citep[][]{Heber+08}. 
}
estimated value of the protosolar nebula is $X\sub{D,PSN}=28.0\pm2.8$~ppm \citep{Asplund+09}, 
and the present-day value of the local interstellar medium varies widely from $X\sub{D,ISM}=13.7\pm5.3$~ppm \citep{Hebrard+05} to at least $32.3 \pm 3.4$ \citep{Linsky+06}.
Classical studies, such as \citet{SST80I}, used a higher value ($X\sub{D}=35~\mathrm{ppm}$) based on classical observations of local interstellar media \citep[e.g.,][]{Vidal-Madjar+Gry84}.
 
Although these values still remain uncertain to some extent and, in particular, $X\sub{D,ISM}$ is still under debate \citep[see][]{Steigman06,Linsky+06,Prantzos+07}, they suggest that the deuterium content evolves with time.
{{Considering that $X\sub{D,PSN}=28$~ppm at 4.57\,Gyr ago, we assume that the deuterium content of present-day star-forming regions may be as low as $20~\mathrm{ppm}$ and use that as our fiducial value.}}
In addition to the time evolution, the deuterium content may show a wide range of values depending on the environment. For example, for stars formed from a cloud strongly affected by winds from nearby evolved stars, we could expect a smaller deuterium content than for other stars. 
It is therefore important to examine how variations in the deuterium fraction affect the PMS evolution of stars. 
In order to do so, we use the {{fiducial}} settings but vary the D/H ratio.

\begin{figure}[!tb]
  \begin{center}
    \includegraphics[width=\hsize,keepaspectratio,clip]{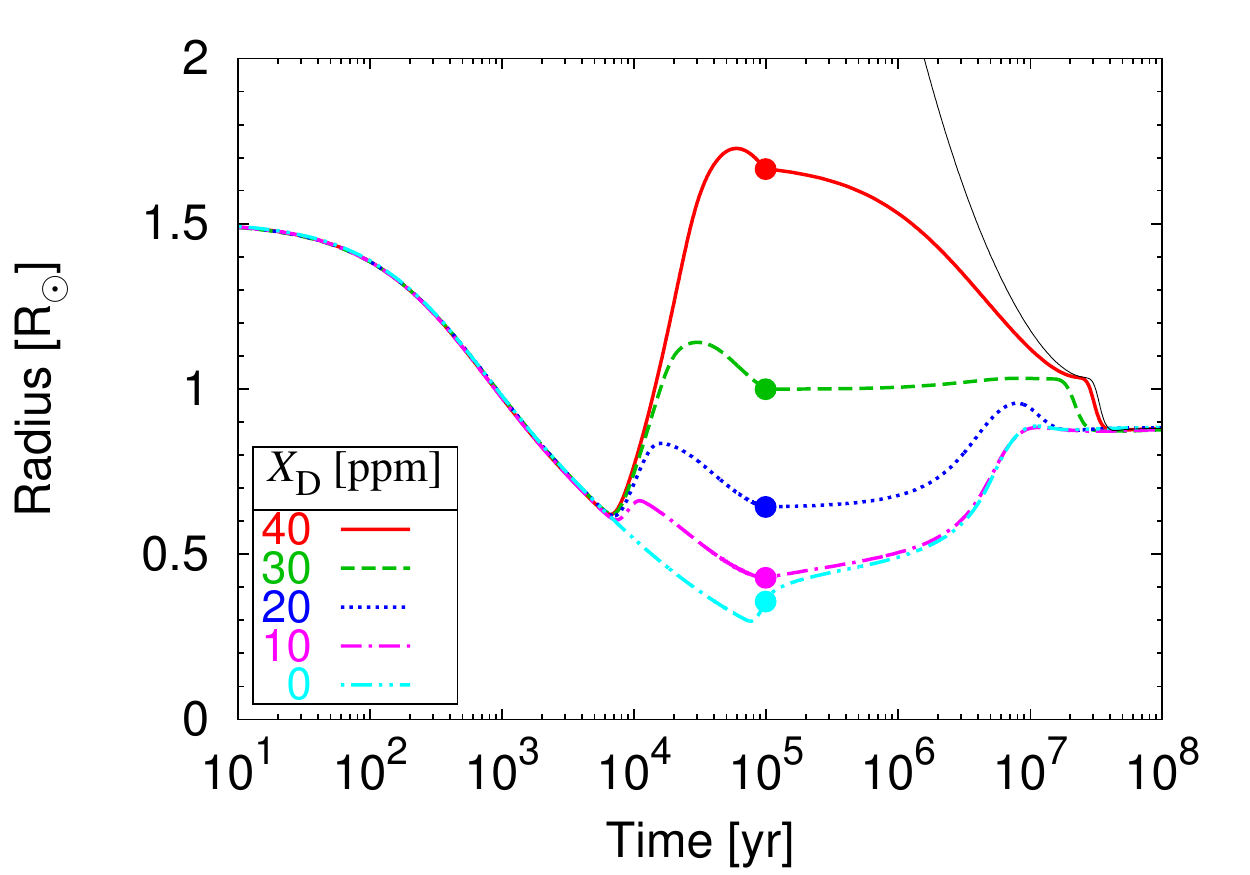}
    \includegraphics[width=\hsize,keepaspectratio,clip]{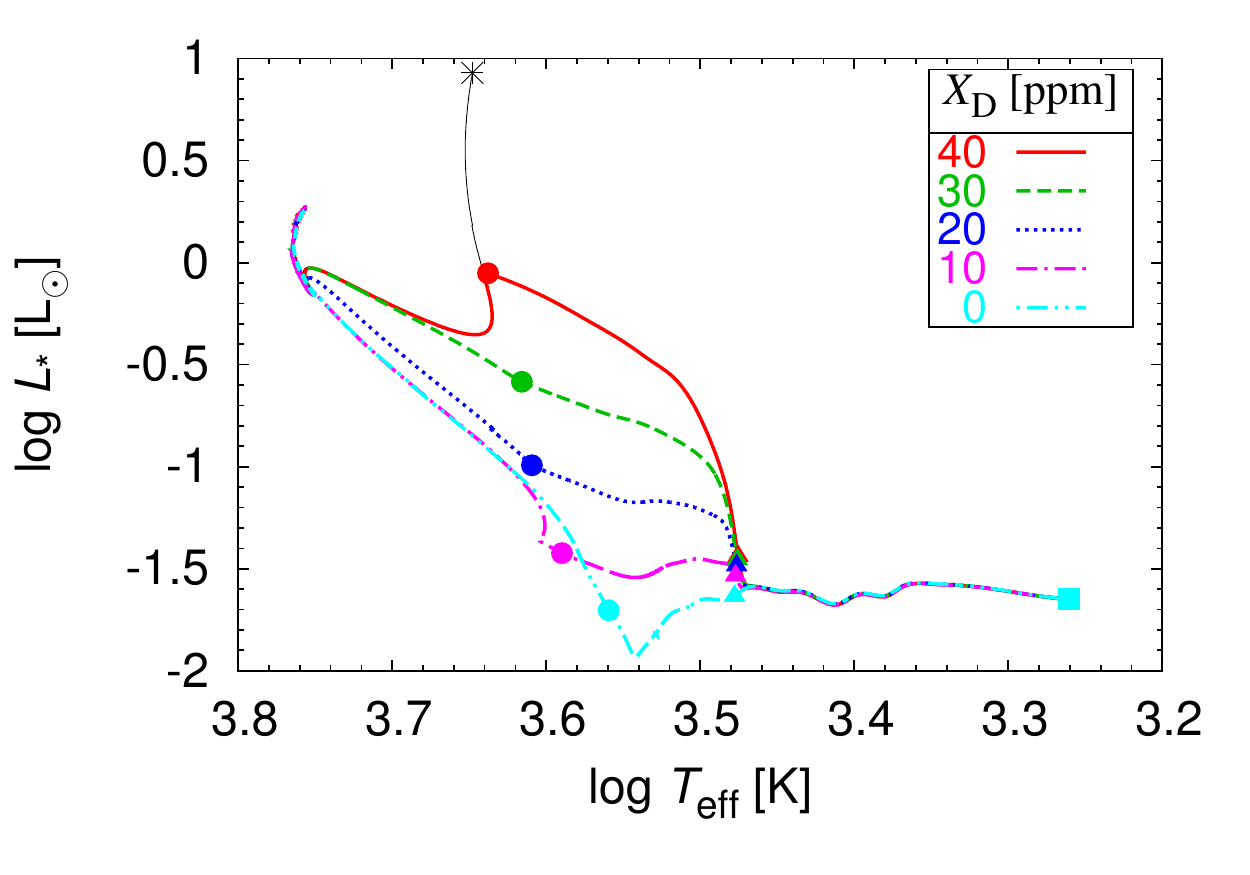}
        \caption{\small{
        Radius evolution (\textit{top panel}) and evolutionary tracks (\textit{bottom}) with varying deuterium contents ranging from $X\sub{D}$ = 40 (red solid line), 30 (green dashed), 20 (blue dotted), 10 (pink dot-dashed), and 0 (blue double dot-dashed)~ppm. The points represent the end of the accretion phase and the black solid lines represent the classical PMS evolutionary tracks.}}
        \label{fig:D}
   \end{center}
\end{figure}

Figure~\ref{fig:D} shows the radius evolution and evolutionary tracks with the different deuterium mass fraction from zero to 40~ppm, where {{the upper limit is set by the cosmic primordial value.}}
{{Whereas}} the evolution before deuterium burning is totally independent of the deuterium content, we {{see that the tracks deviate}} after deuterium fusion sets in. 
For example, in the case $X\sub{D}=40~\mathrm{ppm}$, the radius and luminosity are up to $\sim 4$ and $\sim 30$ times larger, respectively, than for the deuterium-free case.
With the largest deuterium content, $L\sub{nuc}$ exceeds $\frac{1}{7}L\sub{acc}$ for a longer period of time and then the star becomes larger (see Eq.~\ref{eq:lnuccrit2}), {where $L\sub{acc}\equiv GM_{\star}\dot{M}/R_{\star}$ is the gravitational energy of the accreting  material and $L\sub{nuc}$ is the energy production rate by thermonuclear reaction.}
Evolutionary tracks are also largely different depending on $X\sub{D}$ in the region where $\log T\sub{eff}\gtrsim 3.48$, i.e., $T\sub{eff}\gtrsim 3000~\mathrm{K}$.
These large differences illustrate the importance of the deuterium content {{on}} the PMS evolution.

%%%%%%
%%%%%%

In the extreme case $X\sub{D}=0$, the star keeps shrinking even after $\sim 10^4$~years.
Just before the accretion is completed at $10^5$~years, the radius becomes $\sim 0.3~\Rsun$ and eventually the star slightly expands.
This expansion results from hydrogen burning due to sufficiently high central temperature even before the accretion is completed.

We thus find that the deuterium content affects {{low-entropy accretion evolution}} more strongly than {{found by}}
\citet{Stahler88} with spherical accretion. 
This is because the total {{energy injected}} by accretion, $E\sub{acc}=\int L\sub{add} \mathrm{d}t$, exceeds the total energy generated by deuterium burning, 
{{$E\sub{D}=q\sub{D} X\sub{D} M\sub{fin}$, where $q\sub{D}=2.63\times 10^{18}\,\rm erg\,g^{-1}$ is the energy released by deuterium fusion per gram of deuterium}}, when
%%%%%%%%%%%%
\begin{equation}
\xi \gtrsim 0.05 \brafracket{X\sub{D}}{20~\mathrm{ppm}}\brafracket{\bar{R}}{\Rsun}{\brafracket{M\sub{fin}}{\Msun}^{-1}},
\end{equation}
%%%%%%%%%%%%
where $\bar{R}$ is the time-averaged stellar radius.
In spherical accretion, $E\sub{acc}$ always dominates $E\sub{D}$ and the difference in deuterium content is less pronounced. The reverse is true in the case of cold accretion, implying changes in the radii by up to a factor 3 and in luminosity by up to 2 orders of magnitude between the extreme $X\sub{D}$ values.

{
%_______________________________________________
\subsection{Dependence on initial conditions}\label{sec:cold-ini}}
So far we have assumed that our initial stellar seed is characterized by $1.5\,\Rsun$ and $0.01\,\Msun$. As pointed out by \citetalias{Hosokawa+11}, since the entropy of accreting materials is related to the stellar {entropy} in our prescription (see Sect.~\ref{sec:Sacc}), the initial conditions affect the evolution {{\citep[see also][]{Hartmann+97}}}. This is also pointed out by \citetalias{BVC12} who stress the importance of the initial seed mass in the subsequent PMS evolution. Here, for computational reasons, we choose to vary the initial entropy (i.e., radius) instead of the initial mass. As discussed in Appendix~\ref{app:Mini}, the approach is equivalent with the current uncertainties in stellar seed masses leading to values of the radius at $0.01\,\Msun$ between $0.25\,\Rsun$ and $1.5\,\Rsun$.

\begin{figure}[!tb]
  \begin{center}
    \includegraphics[width=\hsize,keepaspectratio,clip]{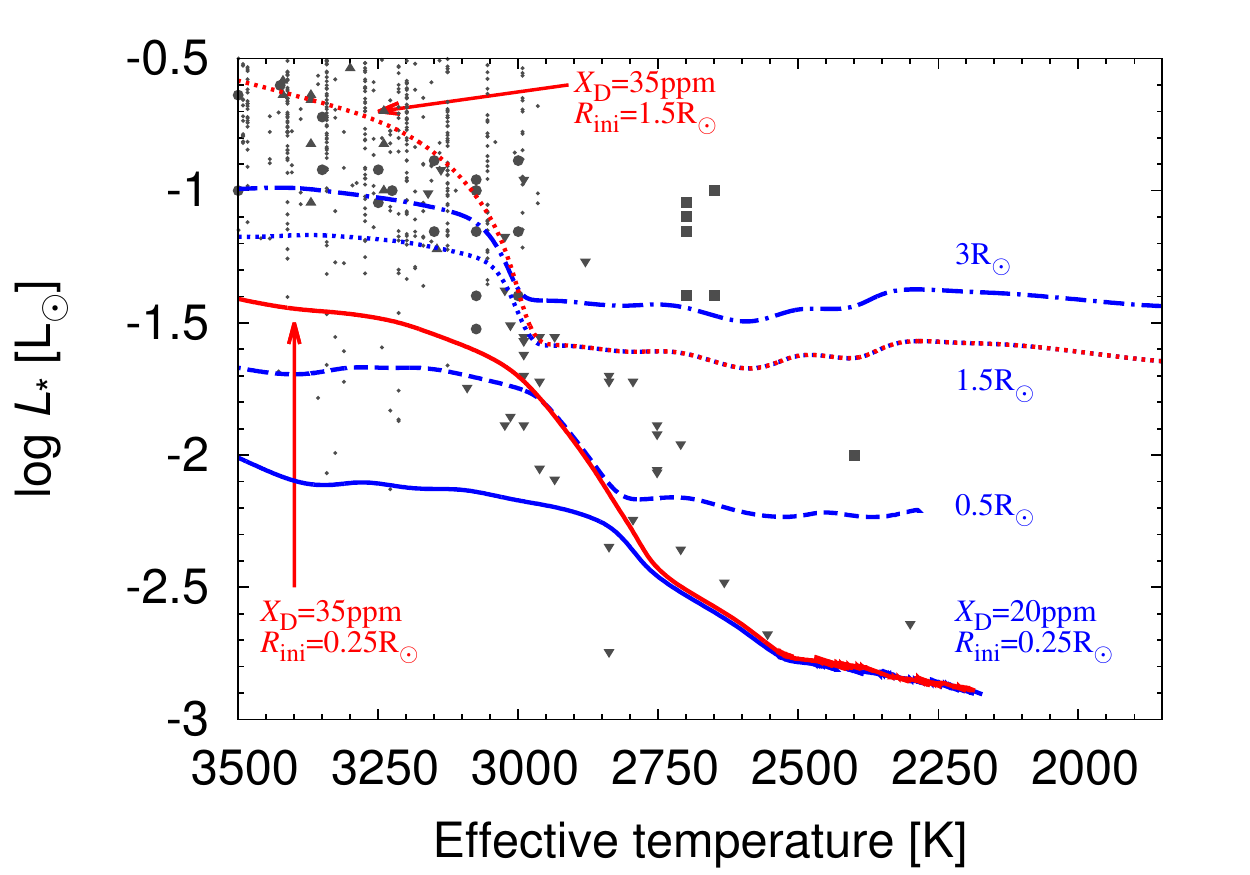}
        \caption{\small{
        Evolutionary tracks {{for different initial radius $R\sub{ini}$ and different initial deuterium content $X\sub{\rm D}$ assuming cold accretion ($\xi=0$).}}
        The solid, dashed, dotted and dot-dashed lines represent the cases $R\sub{ini}=0.25, 0.5, 1.5,$ and $3~\Rsun$, respectively.
        {{The colors indicate deuterium content, $X\sub{\rm D}=20$\,ppm (blue) and $X\sub{\rm D}=35$\,ppm (red).}}
        {{The points represent the observed PMS stars in clusters (see Fig.~\ref{fig:iso-obs}).}}
        {{In all cases, stellar masses reach about $0.5\,\Msun$ at 3500\,K.}}
        }}\label{fig:Rini-iso}
    \end{center}
\end{figure}

In Fig.~\ref{fig:Rini-iso}, we show {the evolutionary tracks of accreting protostars with brown-dwarf masses} obtained with $R\sub{ini}$ from $0.25\,\Rsun$ to $3\,\Rsun$ and $X\sub{D}=20$\,ppm and 35\,ppm. We do not show the results for $R\sub{ini}\gtrsim3\,\Rsun$ since these are similar to the $R\sub{ini}=3\Rsun$ case because of their short K-H timescale.
Like \citetalias{Hosokawa+11} (see their Figs.~5--7), we confirm that different $R\sub{ini}$ values can produce a scatter in the low-temperature region (by up to a 1.5 dex difference in luminosity), which cannot be produced by a varying deuterium abundance.
{{Separately, we note that changing the initial conditions results in {deuterium fusion setting in at a} different stellar mass ($0.08$ and $0.03\,\Msun$ for $R\sub{ini}=1.5$ and $0.25\,\Rsun$, respectively; see Fig.\,\ref{fig:Mini-R}).}}

%%%%%%
%%%%%%

We see that the evolutionary track with $R\sub{ini}=0.25\,\Rsun$ and $X\sub{D}=20$\,ppm is characterized by a very low luminosity.
The main reason for the small luminosity is the small entropy of both an initial seed and accreting materials.
In the cold accretion cases, the entropy of the accreting material is assumed to be the same as the stellar entropy (see Sect.~\ref{sec:Sacc}).
Moreover, the effect of deuterium burning has less impact in this case. 
The star expands in $T\sub{eff}\simeq2500$--$2900\,\rm{K}$ because of deuterium burning and luminosity increases. However the duration of the expansion phase is shorter than the other cases because of the large accreting material's gravitational energy, $L\sub{acc}$\footnote{The luminosity in the present paper is always the stellar intrinsic luminosity and does not include the accretion luminosity emitted from the shock surface.}.
As described in Appendix~\ref{sec:overview}, stars expand when $L\sub{D}>\frac{1}{7}L\sub{acc}$ (Eq.\,\ref{eq:lnuccrit2}). Hence the larger $L\sub{acc}$ makes the expansion phase (phase II in Sect.~\ref{sec:cold}) shorter and the stellar radius and luminosity remain small (see also \citetalias{BVC12}).

%_______________________________________________

\subsection{Evolution tracks for different final masses}\label{sec:Mfin}

\begin{figure}[!tb]
  \begin{center}
    \includegraphics[width=\hsize,keepaspectratio,clip]{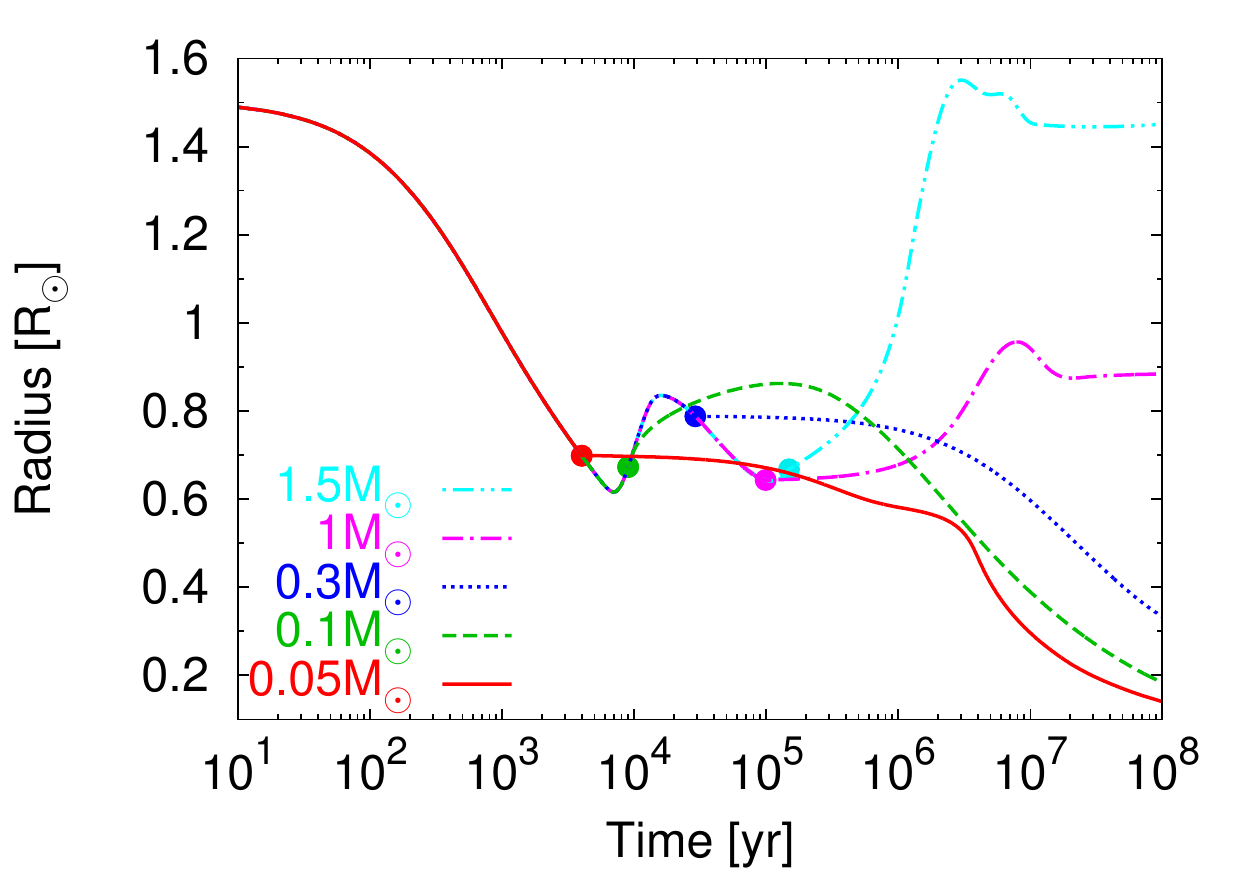}
    \includegraphics[width=\hsize,keepaspectratio,clip]{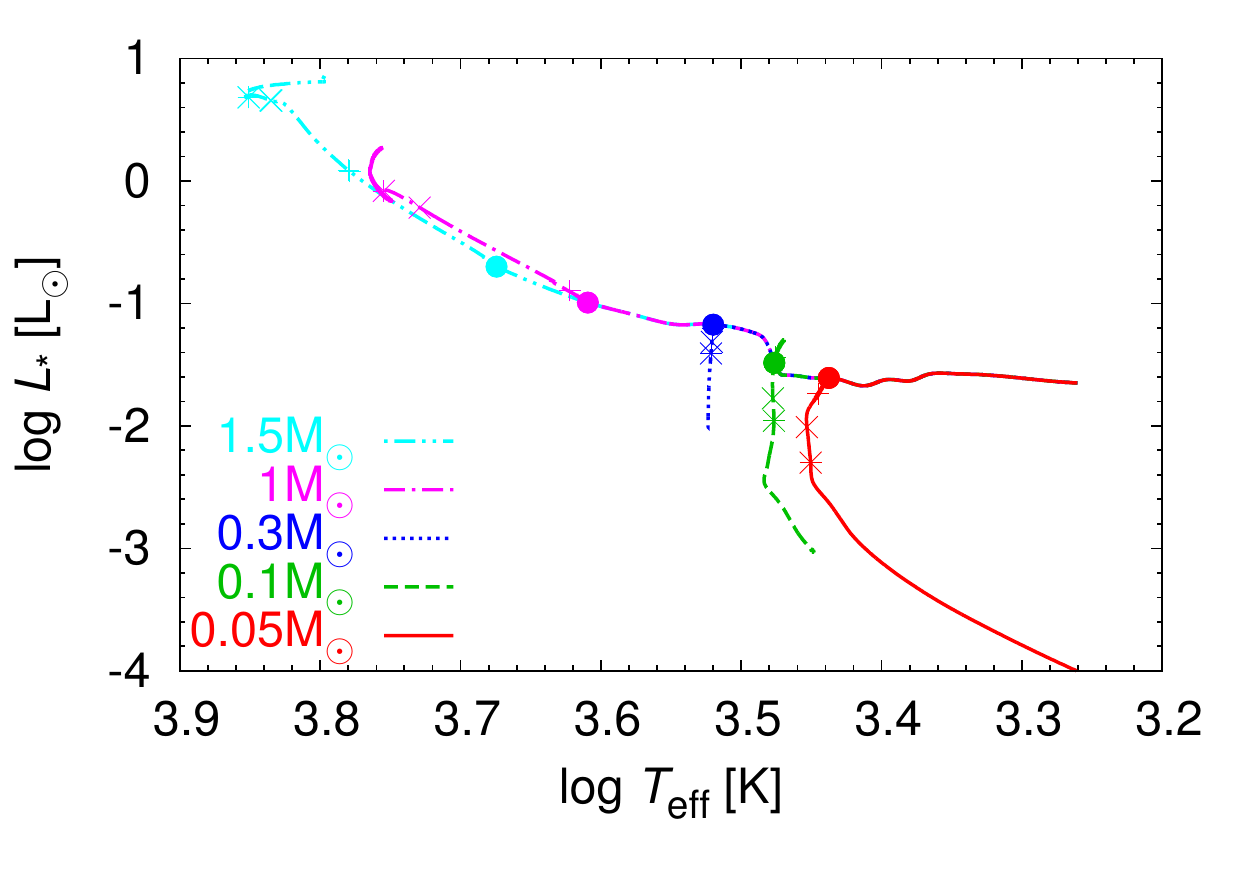}
        \caption{\small{
        Radius evolution (\textit{top panel}) and evolutionary tracks (\textit{bottom}) with varying final masses ranging from $0.05~\Msun$ (solid line), $0.1~\Msun$ (dashed), $0.3~\Msun$ (dotted), $1~\Msun$ (dot-dashed), and $1.5~\Msun$ (double dot-dashed). 
        The accretion is cold ($\xi=0$) and steady ($10^{-5}~\Msun/\textrm{yr}$).
        The filled circles indicate the ages when the accretion is completed and the pluses, crosses, and asterisks represent 1, 5, and 10 million years, respectively.
        The calculation stops at $10^{10}~{\textrm{yr}}$ or when stars start leaving the MS (only in the case of $1.5~\Msun$).
        We note that the evolutionary track of $0.05~\Msun$ is cut short.
        }}\label{fig:Mfin-t-R}
    \end{center}
\end{figure}

In this section we focus on the evolution of stars with varying masses. Apart from the final masses, the settings are the same as the fiducial values listed in Table~\ref{Table:param}.
The cold accretion evolution of stars with masses of 0.05, 0.1, 0.3, 1, and $1.5~\Msun$ are shown in Fig.~\ref{fig:Mfin-t-R}.
Low-mass stars contract for a long time because of their long K-H timescale. For example, stars with 0.1 and $0.3~\Msun$ enter their MS at $10^9$ and $3 \times 10^8$ years, respectively. During the first million years of evolution, Fig.\,\ref{fig:Mfin-t-R} shows that the size of stars undergoing cold accretion is not a monotonic function of mass. For example, a $0.1~\Msun$ star expands just after accretion ceases due to deuterium burning and becomes larger than the other stars considered here  for some
time.  On the other hand, a $0.05~\Msun$ star contracts {{monotonically}} and even more after $\simeq 1$--2 million years when all deuterium has been consumed. 
Higher mass stars ($1.0$, $1.5~\Msun$) expand just after accretion has stopped because of hydrogen burning.

%%%%%%
%%%%%%

%%______________________________
\subsection{Comparison with previous studies} \label{sec:comp_prev}

Here we compare our results with the studies of \citetalias{BC10} and \citetalias{Hosokawa+11}. We use two particular examples: the long-dashed line in Fig.~2 of \citetalias{BC10} (episodic and cold accretion) and the case of ``mC5-C'' in \citetalias{Hosokawa+11} (steady and cold accretion). The other settings are summarized in Table~\ref{Table:param}.
{{Quantitatively, their results differ: {\citetalias{Hosokawa+11} obtain radii that are often} about two times larger than those of \citetalias{BC10}. {For example, in} \citetalias{Hosokawa+11}, the radius when $M_\star = 0.9~\Msun$ is $\sim 1.3~\Rsun$ (see their Fig.~2) {while for} the same mass, it is $\sim 0.6~\Rsun$ in \citetalias{BC10}.}}

\begin{figure}[!tb]
  \begin{center}
    \includegraphics[width=\hsize,keepaspectratio,clip]{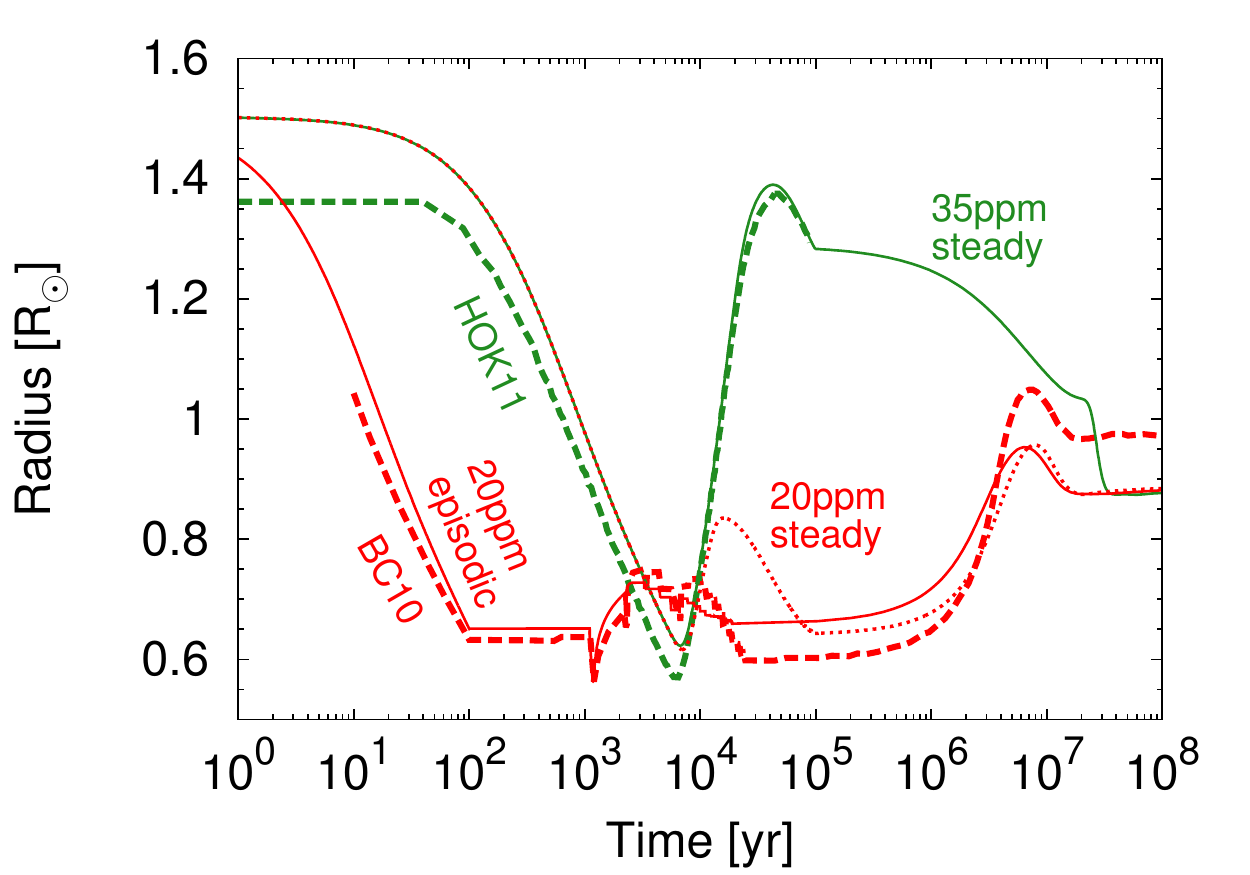}
        \caption{\small{
            {{Cold-accretion ($\xi=0$)}} evolution of the stellar radius with time for the studies of \citetalias{Hosokawa+11} (dashed green) and \citetalias{BC10} (dashed red) and for our results with steady accretion and $X\sub{D}=35\,\mathrm{ppm}$ (plain green), $X\sub{D}=20\,\mathrm{ppm}$ (dotted red). A model with episodic accretion (as described in Sect.~\ref{sec:episodic}) and $X\sub{D}=20\,\mathrm{ppm}$ is also shown (plain red). 
        }}
        \label{fig:comp_prev}
   \end{center}
\end{figure}

An important difference {{between these calculations}} is that the assumed deuterium contents differ by a factor 1.75.  As shown in the previous section, this difference has a large impact on the evolution (see Fig.~\ref{fig:D}). 
Figure~\ref{fig:comp_prev} compares the radius evolution in \citetalias{BC10} and \citetalias{Hosokawa+11} to our calculations with the same hypotheses. Given the fact that different stellar models are used (including different values of $\alpha\sub{MLT}$ --see Table~\ref{Table:param}), there is excellent agreement between our calculations and those of \citetalias{BC10} and \citetalias{Hosokawa+11}. The difference between the former and the latter is indeed caused by the different assumed deuterium abundances, $X\sub{D}=20~\mathrm{ppm}$ and $35~\mathrm{ppm}$, respectively. 
{The figure also compares two calculations with episodic and steady accretion for $X\sub{D}=20~\mathrm{ppm}$. The difference between these two calculations is much smaller than between calculations with different deuterium abundances: the evolution of the PMS star is principally governed by the {\textit{mean}} accretion rate and by the deuterium abundance of accreted material and much less by the nature (steady or episodic) of the accretion.}

%%%%%%
%%%%%%

%______________________________
\section{Hot and warm accretion} \label{sec:hot}

In this section we explore {{how heat injection by accretion affects stellar evolution. First, we assume that accretion energy is redistributed uniformly in the star and examine variations with the accretion efficiency parameter $\xi$ defined in Eq.~\eqref{eq:Ladd}. Then, we examine, with the parameter $m\sub{ke}$ (see Eq.~\ref{eq:eps_linear}), how assumptions concerning where the accretion energy is released affect the PMS evolution tracks.}}
In this section the parameters other than $\xi$ and $m\sub{ke}$ are set to their default values (see Table~\ref{Table:param}).

\subsection{Early evolution with a varying $\xi$}\label{sec:xi}

\begin{figure}[!tb]
  \begin{center}
    \includegraphics[width=\hsize,keepaspectratio,clip]{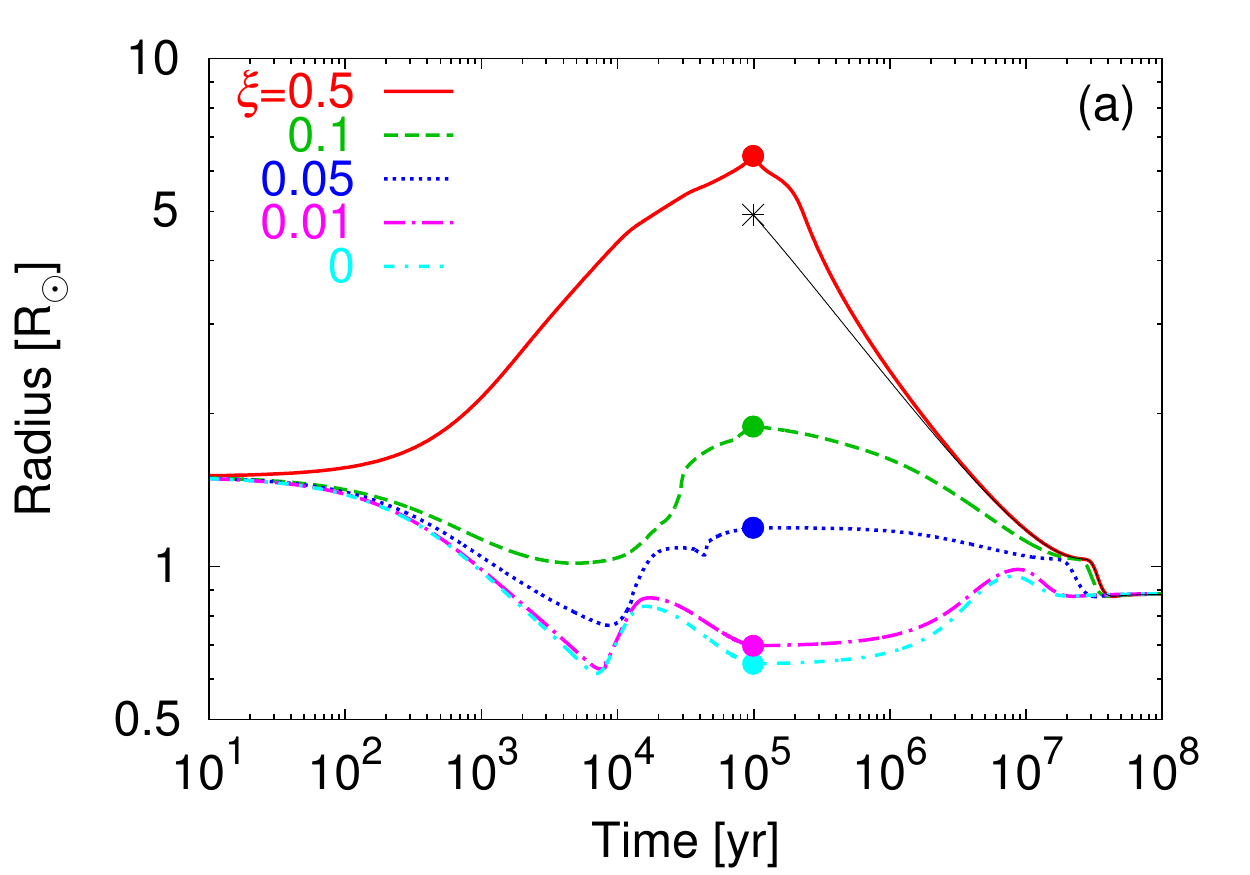}
    \includegraphics[width=\hsize,keepaspectratio,clip]{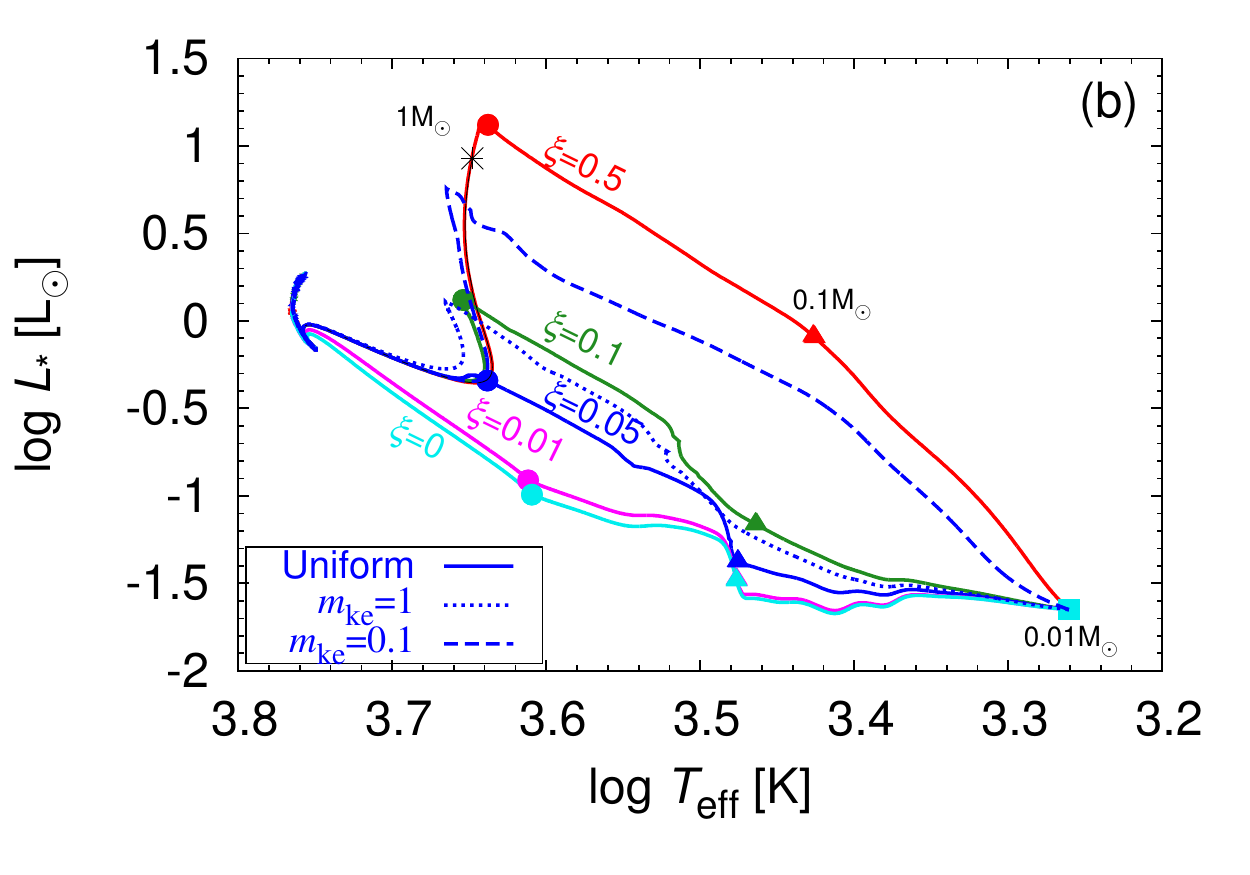}
        \caption{\small{
        \textit{Top panel.} 
        Radius evolution for various heat injection efficiencies, $\xi=0.5$ (solid line), 0.1 (dashed), 0.05 (dotted), 0.01 (dot-dashed), and 0 (double dot-dashed). 
        The classical PMS evolution, which is indicated by the thin solid line and starts at the asterisks, has a substantial overlap with the lines of $\xi=0.5$.
        \textit{Bottom panel.}
        Evolutionary tracks with different $\xi$ ranging 0.5 (red line), 0.1 (green), 0.05 (blue), 0.01 (magenta) and 0 (cyan).
        The solid, dotted, and dashed lines represent the cases with uniform heat injection, $m\sub{ke}=1,$ and $m\sub{ke}=0.1$, respectively.
        In both panels, we adopt $X\sub{D}=20\,\rm{ppm}$ and $R\sub{ini}=1.5\,\Rsun$.
        }}\label{fig:hot}\label{fig:x}
    \end{center}
\end{figure}

Figure~\ref{fig:hot} shows {{the evolution tracks}} obtained with the heat injection parameter in the range $\xi=0$--0.5.
When comparing simulations with $\xi=0.5$ and $\xi=0$, the radii differ by 1 order of magnitude and the luminosities differ by 2 orders of magnitude.
{{These differences in evolutionary tracks are {larger and concern a wider} range of effective temperatures {than those due to a} different deuterium content.
Thus, the most important parameter controlling the PMS evolution is the amount of heat injected by accretion. }}

The evolutionary track of accreting stars corresponds to the ``birthline'' proposed by \citet{Stahler+83}.
As \citetalias{BVC12} pointed out, Fig.~\ref{fig:hot}b shows that the birthlines strongly depend on heat injection $\xi$, implying that the concept of a definitive birthline may be elusive. The radius when the accretion is completed (i.e., at $10^5$~yr) in the case of $\xi=0.5$ is almost the same as obtained by \citet{Stahler+Palla05} for spherical accretion.

%%%%%%
%%%%%%

Figure~\ref{fig:hot}a  shows that before deuterium burning (at $10^4$ to $10^5$\,yrs), the radius initially increases for $\xi=0.5$, decreases slightly for $\xi=0.1,$ and shows a pronounced decrease for $\xi\lesssim 0.05$. 
Again, this behavior can be understood using the energy equation {as in Appendix\,\ref{sec:overview}}.
The situation corresponds to a case when nuclear reactions and radiative cooling may be neglected (i.e., $L\sub{nuc}, L_\star \ll L\sub{acc}$). The energy equation may thus be shown to yield
%%%
\begin{equation}
\frac{\dot{R}}{R_\star} = \left( 2-\frac{1-\xi}{C}\right) \frac{\dot{M}}{M_\star}\, \label{eq:hot_energy}
,\end{equation}
%%%
where $C$ is a constant as a function of the specific heat ratio and the polytropic index (see Appendix~\ref{app:anal}).
Therefore, the radius should be constant if $\xi = 1- 2C$.
In the case of a fully convective star with monoatomic gas ($C=3/7$), the critical $\xi$ for the constant radius is $1/7$.

In a second phase, for $\xi\lesssim 0.1$, deuterium burning sets in and leads to a limited increase of the stellar radius. However, this does not occur in the case of $\xi=0.5$. First, in this case, deuterium fusion starts only at $\sim 1.2\times10^5$~years, i.e., after the accretion is completed because of the low central temperature during the accreting phase due to the large radius.
Moreover, since $L\sub{nuc}$ does not exceed $L_\star$, deuterium burning only delays the contraction from radiative cooling, and the star does not expand (see Appendix~\ref{app:anal} and Eq.~\ref{eq:KHMacc}).

%________________________________________________________
\subsection{Importance of the location of heat injection}\label{sec:mke}
So far, we assumed a uniform distribution of injected heat as in \citetalias{BC10} (see Eq.~\ref{eq:eps_uniform}).
However, this assumption may not {{be}} valid, especially in stars with a large radiative core.
Moreover, as described in Sect.~\ref{sec:Sacc}, recent radiation-hydrodynamic simulations by \cite{Geroux+16} showed that the uniform heat redistribution may not be accurate.
We now examine the case in which the accretion heat is injected only in surface regions (see Eq.~\ref{eq:eps_linear}).

{{Figure~\ref{fig:x}b}} shows that the evolution of an accreting solar-mass star strongly depends on the assumed location of the heat injection region. 
{{In the case of $\xi=0.05$ (blue lines), the luminosity with $m\sub{ke}=0.1$ is up to about one order of magnitude larger than the case of the uniform distribution and is almost the same as the case with $\xi=0.3$ and uniform distribution. }}
This dependence mainly comes from the assumption that the entropy of accreting materials are the same as the stellar surface (see Sect.~\ref{sec:Sacc}).
If $m\sub{ke}$ is small and $\xi>0$, the accretion heat is injected only in the outer envelope. This causes $\varepsilon\sub{add}$ to be large at these locations, which, according to Eq.~\eqref{eq:dLdM}, leads to an increase of the stellar luminosity and hence of the specific entropy in this region. The surface entropy then becomes larger than in the case of a large $m\sub{ke}$, or equivalently, of a uniform injection of accretion heat. The assumption that any added mass has the same entropy as the stellar photosphere effectively leads to accreting material with a higher entropy and therefore to a larger radius. 

The assumptions of a uniform or linear injection of accretion energy and of continuity of entropy between the photosphere and accreted material are questionable. However, we can see that the uniform model with low $\xi$ effectively represents one low extreme model. For high $\xi$ values ($\xi\wig{>}0.3$), we expect evolution models to evolve very rapidly initially, losing memory of the initial conditions and resembling standard evolutionary tracks. In that sense, although radiation hydrodynamic simulations would be desirable, the uniform model defined by Eq.~\eqref{eq:eps_uniform} is a useful simplification to represent possible evolutionary tracks.

%________________________________________________________
\section{Implication for the evolutionary tracks in the H-R diagram}\label{sec:HR}

In this section we compare our results with observations in the H-R diagrams.
We now turn to isochrones by integrating the evolutions of various final masses as shown in Sect.~\ref{sec:Mfin}.
We give special attention to whether these new evolutionary tracks can explain the \textit{luminosity spread problem} for young stellar objects (YSOs), i.e., the fact that for a given cluster, stars are spread over a relatively wide range of luminosities instead of forming a well-defined luminosity-effective temperature relation as would be expected for stars of similar ages and compositions. 
Although the consequences of accretion in the H-R diagrams have been investigated by previous works (\citetalias{BCG09}, \citetalias{Hosokawa+11} and \citetalias{BVC12}), we choose here to attempt to constrain the values of $\xi$ that are in agreement with the observations of young clusters.

%________________________________________________________

\subsection{The PMS luminosity spread problem}\label{sec:iso}

\subsubsection{Observational constraints}\label{sec:lspread}
The luminosity spread of PMS stars has been a matter of debate for decades \citep[see the review of][]{Hillenbrand09,Jeffries12,Soderblom+14}.
This spread is seen almost ubiquitously in star-forming regions and young clusters. 
Three types of explanations have been proposed: (i) observational or astrophysical uncertainties \citep[e.g.,][]{Hartmann01}, (ii) an intrinsic age spread \citep[e.g.,][]{Palla+Stahler00,Inutsuka+15} and (iii) physical processes \citep[e.g.,][]{Chabrier+07,BCG09}.
Determining the reason for this spread is important for our understanding of star formation.

It is difficult to determine the luminosity of young PMS stars because it is subject to the observational (e.g., differential extinction, reddening, distance, and cluster membership) and astrophysical (e.g., circumstellar material and its accretion, unresolved binary and variability) uncertainties \citep[e.g.,][]{Hartmann01}.
However, the contribution to the luminosity spread by each uncertainty has been quantitatively estimated \citep[e.g.,][]{Reggiani+11,Burningham+05} and various authors claimed that the sum of their contributions is smaller than the observed luminosity spread. Moreover, \citet{Jeffries07} found that the projected radii, which are less affected by observational uncertainties, instead of luminosities, also spread widely.
These results suggest that the luminosity spread is genuine.

A luminosity spread of 0.2--0.3\,dex would correspond to a $\sim 0.4$\,dex spread in the ages of PMS stars. This could be explained theoretically; for example, \citet{Inutsuka+15} propose that stars are formed by the recurrent compressions by expanding bubbles and that consequently the members of a cluster are not necessarily formed in a short period of time. 

Instead, cold accretion leads to much smaller stellar radii and luminosities than the classical, non-accreting models and this can explain the luminosity spread for stars $T\sub{eff}\gtrsim 3500~\rm{K}$ {(see \citetalias{BCG09,Hosokawa+11,BVC12};  see also Sect.~\ref{sec:comp_classical})}. For stars of lower temperatures, the results depend on the size of the assumed seed radius and differ between the different studies (see Sect.\,\ref{sec:cold-ini}).

{{
After showing {how the PMS isochrones depend on $\xi$, $X\sub{D}$, and $R\sub{ini}$,} we try to estimate possible values of $\xi$ that are compatible with the observations. In this section, {we assume a uniform injection of accretion heat} (see Sect.\,\ref{sec:short-conclusion}).
}}

%________________________________________________________

\subsubsection{Isochrones as a function of assumed $\xi$} \label{sec:hot-HRD}

\begin{figure}[!tb]
  \begin{center}
    \includegraphics[width=\hsize,keepaspectratio,clip]{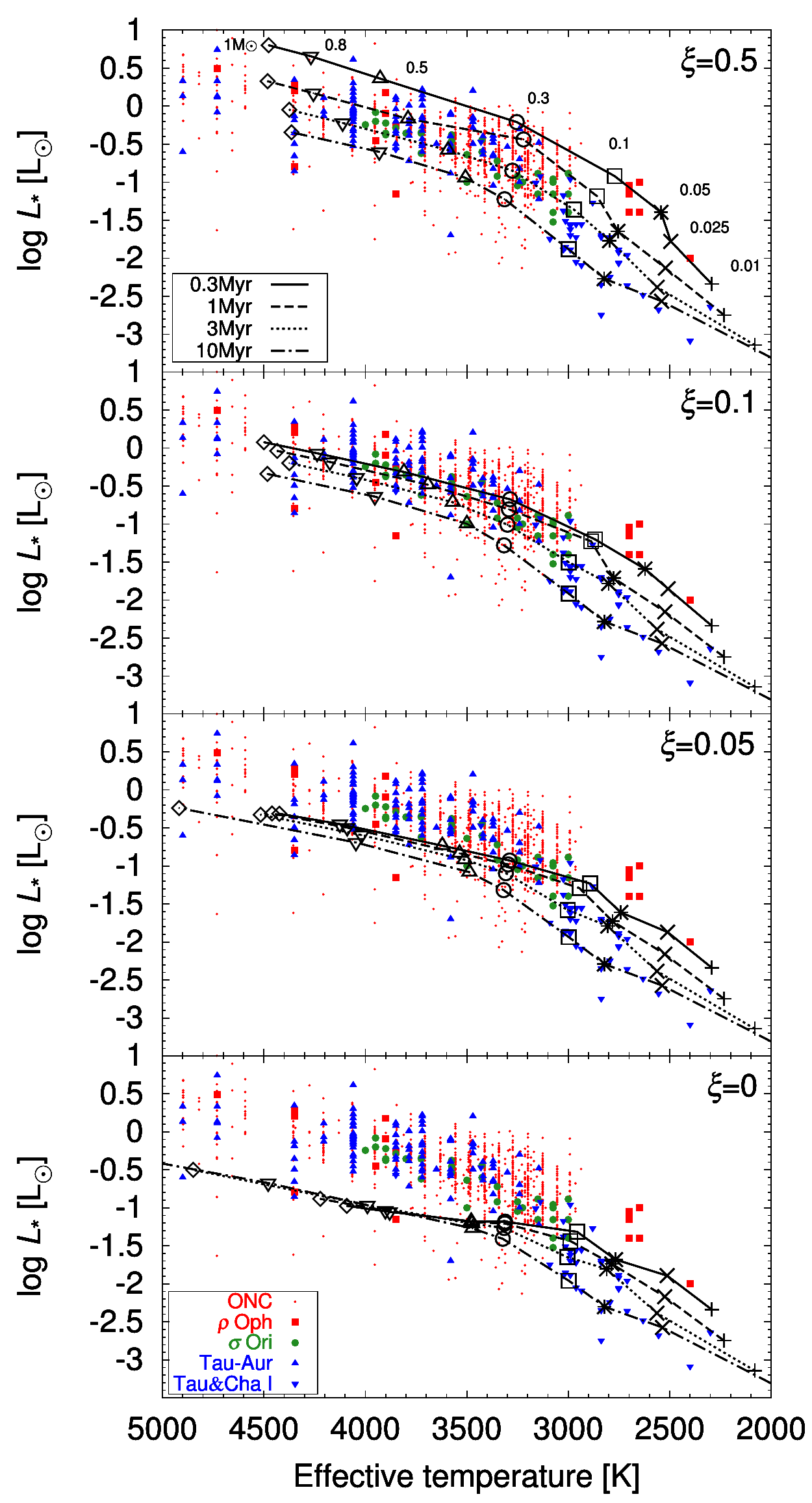}
        \caption{\small{
        Comparison of 0.3 (solid line), 1 (dashed), 3 (dotted), and 10 (dot-dashed) million-year isochrones for the cases $\xi=0.5$, 0.1, 0.05, and 0 from top to bottom {{in the H-R diagram}} with observed PMS stars:
        {{
        ONC (red, small diamonds), $\rho$ Oph (red squares), $\sigma$ Ori (green circles), Tau-Aur (blue triangles), and Tau and Cha I (blue, inverted triangles).
        }}
        Each isochrone is comprised of evolutionary tracks of 0.01, 0.25, 0.05, 0.1, 0.3, 0.5, 0.8, and 1~$\Msun$ stars denoted by pluses, crosses, asterisks, open squares, open circles, open triangles, open inverted triangles, and open tilted squares, respectively.
        The initial radius is $1.5~\Rsun$ and $X\sub{D}=20\,\rm{ppm}$.
        The isochrones with varying $\xi$ cover the wide range of the region where the observed stars exist in the high-temperature regime.
        }}\label{fig:iso-obs}
    \end{center}
\end{figure}

In Fig.~\ref{fig:iso-obs} we compare the PMS stars observed in several young clusters to our theoretical isochrones for the cases $\xi=0$, $0.05$, $0.1,$ and $0.5$. 
The remaining parameters are our fiducial values ($X\sub{D}=20$\,ppm, $R\sub{ini}=1.5\,\Rsun$, and $M\sub{ini}=0.01\,\Msun$).
We use the following observational data of young stars: $\rho$ Ophiuchus \citep{Gatti+06}, $\sigma$~Orionis \citep{Gatti+08}, Taurus and Chamaeleon I \citep{Muzerolle+05}, Taurus-Auriga \citep{Kenyon+Hartmann95}, and Orion nebula cluster \citep[ONC;][]{Da-Rio+10}.

%%%%%%
%%%%%%

The first panel shows that classical evolutionary tracks with $\xi=0.5$ would require a large spread of ages to explain all clusters. The ages of stars would need to range between 0.3 and 10\,Myr in {$\rho$ Oph, ONC, and Tau-Aur} and between 1 and 10\,Myr in the other clusters to reproduce the observed luminosity spread. Even in that case, a few stars are underluminous and can be explained only by invoking that their luminosity has been underestimated. \citet{Gatti+06} estimated that the errors in the luminosities of these stars is $\sim0.5$~dex, which indeed means that this is a possibility. 

Other fixed values of $\xi$ do not allow us to find a better solution. When $\xi=0.1$, the most luminous stars cannot be explained anymore. When $\xi\le 0.05$ the slope of the isochrones becomes inconsistent with the ensemble of observational data points. At the same time,  the $\xi=0$ isochrones are characterized by very low luminosities and can explain the stars with the lowest luminosities observed in {$\rho$~Oph}. 

Conversely, if one assumes that stars within a cluster are coeval, a distribution of the values of $\xi$ within a cluster can be invoked to explain the luminosity spread, as proposed by \citetalias{BCG09}, \citetalias{Hosokawa+11}, and \citetalias{BVC12}.
However, this approach fails to reproduce the luminosity spread of very-low-mass stars in the {Tau and Cha I}. We revisit this in Sect.~\ref{sec:ini}.

%________________________________________________________
\subsubsection{Dependence on deuterium content} \label{sec:D-iso}

As described in Sect.~\ref{sec:coldD}, even the deuterium fraction of the present-day local ISM is still under debate.
{{We now investigate {how isochrones depend on} deuterium content.}}

\begin{figure}[!tb]
  \begin{center}
    \includegraphics[width=\hsize,keepaspectratio,clip]{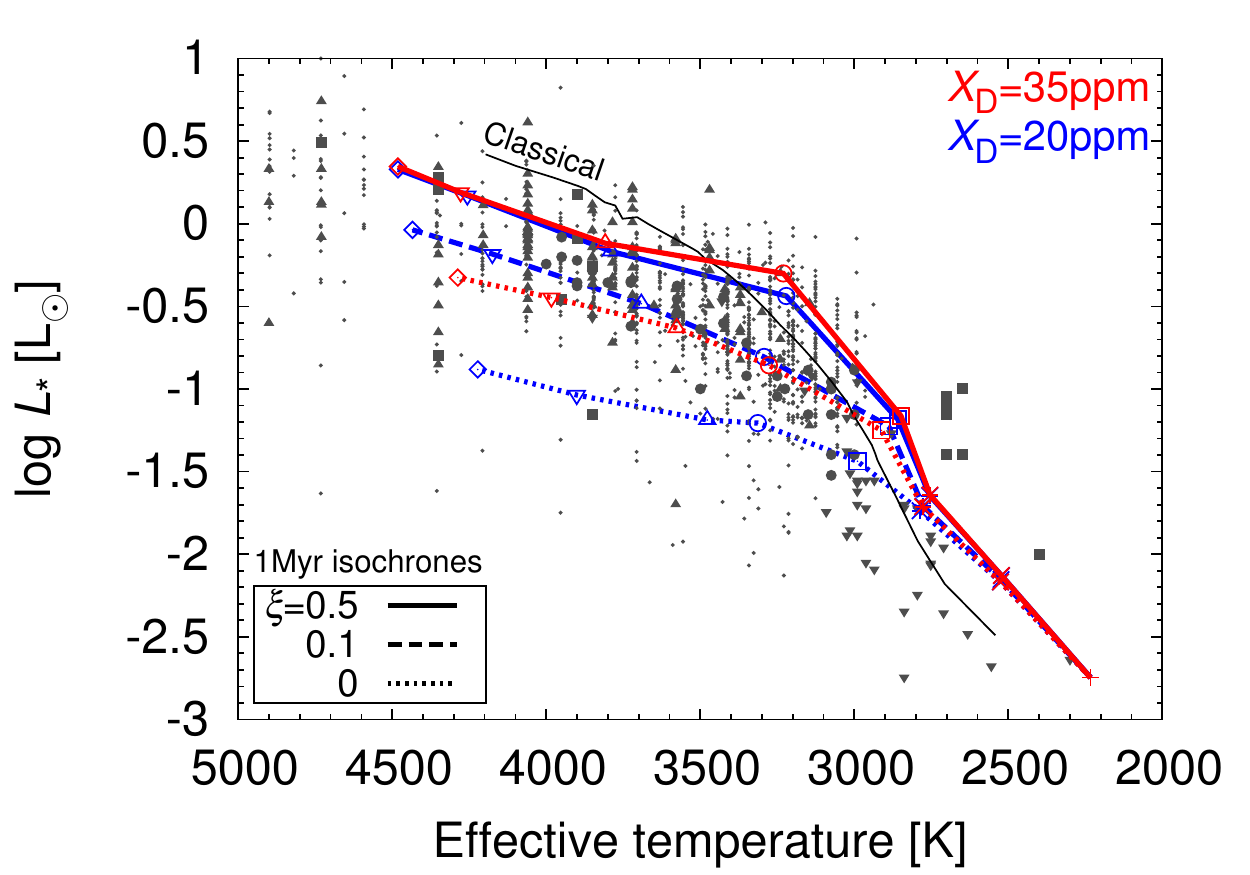}
        \caption{\small{
Comparison of 1~Myr isochrones for the cases $X\sub{D}=35$ ({{red lines}}) and 20~ppm ({{blue}}) ranging $\xi$ from 0 (dotted) to 0.5 (solid), respectively. The initial radius is $1.5~\Rsun$.
        {{The points for observation and isochrones are the same as in Fig.~\ref{fig:iso-obs}.
        }}
        The isochrone with $(X\sub{D},\xi)=(35~{\rm{ppm}}, 0)$ is almost the same as that with (20~ppm, 0.1), implying that the different deuterium content has a large impact on the PMS evolution.
        The thin solid line is a 1~Myr non-accreting isochrone \citep{Baraffe+98}. 
        }}\label{fig:D-iso}
    \end{center}
\end{figure}

Figure~\ref{fig:D-iso} shows 1\,Myr isochrones obtained with different values of $\xi$ and with either $X\sub{D}=20$\,ppm or $35$\,ppm. The isochrones obtained $\xi=0.5$ are almost independent of $X\sub{D}$. However, isochrones obtained for $\xi=0$ differ very significantly when $X\sub{D}$ changes. As described in 
{{Sect.~\ref{sec:typical}}}, this is because in the case of cold accretion, deuterium burning regulates the PMS evolution.
But Fig.~\ref{fig:D-iso} also shows that the isochrones remain almost parallel. For example, the isochrone with $(\xi, X\sub{D})=(0.1, 20~{\rm ppm})$ is very similar to that with $(0, 35~{\rm ppm})$.
This is because both $\xi$ and $X\sub{D}$ control the specific entropy of the accreted material.

In the context of a variable $\xi$ value, a low abundance of deuterium yields a more important spread of solutions than a high abundance. We do not expect $X\sub{D}$ to vary within a cloud because of turbulent mixing. However, if $X\sub{D}$ varies from one cluster to the next, this would yield a more important luminosity spread for clusters with low $X\sub{D}$ values. 

%%%%%%
%%%%%%

%________________________________________________________

\subsubsection{Dependence on initial conditions} \label{sec:ini}

\begin{figure}[!tb]
  \begin{center}
    \includegraphics[width=\hsize,keepaspectratio,clip]{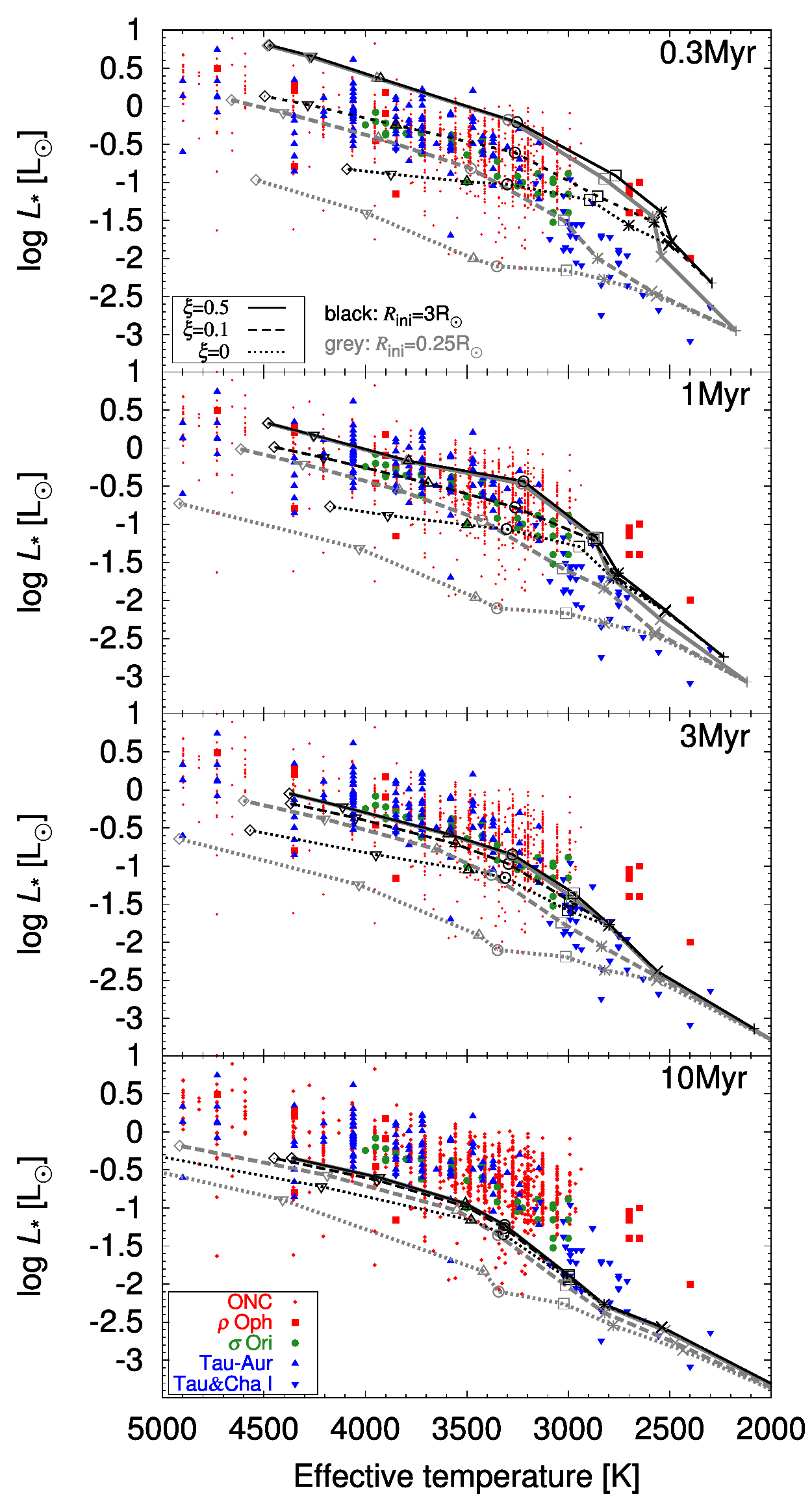}
        \caption{\small{
        Comparison of isochrones at 0.3, 1, 3, and 10 {{Myr}} from top to bottom.
        {{
      $R\sub{ini}$ is either $3\,\Rsun$ (gray lines) or $0.25\,\Rsun$ (black).
        The solid, dashed, and dotted lines represent isochrones with $\xi=0.5, 0.1,$ and $0$, respectively.
        }}
        Here $X\sub{D}=20\,\rm{ppm}$.
        {{The points are the same as in Fig.~\ref{fig:iso-obs}.
        }}
        }}\label{fig:iso-ini}
    \end{center}
\end{figure}

As described in Sect.~\ref{sec:cold-ini} {in the context of cold accretion, the PMS evolutionary tracks depend on the initial conditions, and in particular on the physical characteristics of the initial stellar seed.} 
In Fig.~\ref{fig:iso-ini}, we show the isochrones obtained with $R\sub{ini}$ {of} $0.25\,\Rsun$ {and} $3\,\Rsun$, for different values of $\xi$. 
The comparison of the results with the same $\xi$ and different $R\sub{ini}$ shows that the dependence on $R\sub{ini}$ is significant across the entire effective temperature range, except for hot accretion ($\xi\sim 0.5$). The dependence is most important for cold accretion (see dotted lines in Fig.~\ref{fig:iso-ini}) and corresponds to the situation described in Sect.~\ref{sec:cold-ini}. For intermediate values $\xi\sim 0.1$ (dotted lines), the dependence on $R\sub{ini}$ is smaller but still significant. It is most pronounced at small ages ($\lesssim1\,\rm{Myr}$, which corresponds to the K-H timescale) and for low-mass stars; since the total energy of high-mass stars is dominated by the injected energy, $E\sub{add}$, the initial energy does not matter.

%%%%%%
%%%%%%

A high value of $R\sub{ini}\wig{>}1.5~\Rsun$ is required to explain the most luminous stars,
particularly for the low $T\sub{eff} <3000\,$K stars.
However, invoking a variable $R\sub{ini}$ including low values $R\sub{ini}\sim 0.25\,\Rsun$ is required to explain the observed luminosity spread in the {Tau and Cha~I} clusters by assuming coeval stars.

%________________________________________________________
\subsubsection{Resulting constraints} \label{sec:short-conclusion}

{
The observed luminosities and effective temperatures of young clusters can hence be explained in the context of coeval stars by assuming that $\xi$ and $R\sub{ini}$ can vary within a given cluster. 
This interpretation suggests that $\rho$ Oph, ONC, and Tau-Aur}
{{are extremely young ($\sim0.3\,\rm{Myr}$); $\sigma$ Ori is slightly older at $\sim 1\,$Myr\footnote{
{We do not derive the age of Tau and Cha~I because the data in \citet{Muzerolle+05} is an assembly of young stars in two star-forming regions.}}.
However, inaccurate estimations of stellar luminosities, membership issues, and other problems can affect these age estimates. }}

{{We can see in Sect.~\ref{sec:hot-HRD} and Fig.~\ref{fig:iso-obs} that for a majority of stars $\xi\wig{>}0.1$ for $X\sub{D}=20$\,ppm and $R\sub{ini}=1.5\,\Rsun$ considering the following two facts: the slopes of the isochrones are not compatible with the observations for $\xi\le 0.05$ and the number of underluminous stars is small in $T\sub{eff}=3500$--4500\,K.
For higher values of $X\sub{D}$, this constraint on $\xi$ should be relaxed somewhat. 
For lower values of $R\sub{ini}$, we would have to impose $\xi$ to be even closer to $0.5$. 
Even with higher values of $R\sub{ini}$, $\xi\wig{<}0.1$ would still be rare because the slope of the isochrone does not match the observation.
It is thus premature at this point and without a proper modeling of where the accretion heat is deposited to attempt constraining $\xi$ more precisely. 
However, one important conclusion that we can draw is that the evolution tracks cannot deviate too significantly from the standard model: stars with entropies that are too low are ruled out by the observation of young clusters. In that sense, the limit set by the model characterized by a uniform accretion heat redistribution, $\xi=0.1$, $X\sub{D}=20$\,ppm and $R\sub{ini}=1.5\,\Rsun$, is useful when considering the range of possibilities in agreement with observational data. 
}}

{{
Finally, if indeed star growth is characterized by an inefficient burial of accretion heat ($\xi < 0.5$), then we should expect that clusters characterized by a lower deuterium mixing ratio $X\sub{D}$ should have a larger luminosity spread than clusters with a higher $X\sub{D}$. This may thus become testable, depending on the possibility to reliably determine D/H ratios in clusters (see Sect.~\ref{sec:coldD}). 
}}

%________________________________________________________
\subsection{The case of CoRoT 223992193}

We now consider the case of a specific system, the eclipsing binary CoRoT 223992193 {{\citep{Gillen+14,Stassun+14}}}. 
We expect both stars to have the same age. Therefore, the system provides important constraints to test our evolution models and retrieve $\xi$ independent of the cluster results discussed previously. 

\begin{figure}[!tb]
  \begin{center}
    \includegraphics[width=\hsize,keepaspectratio,clip]{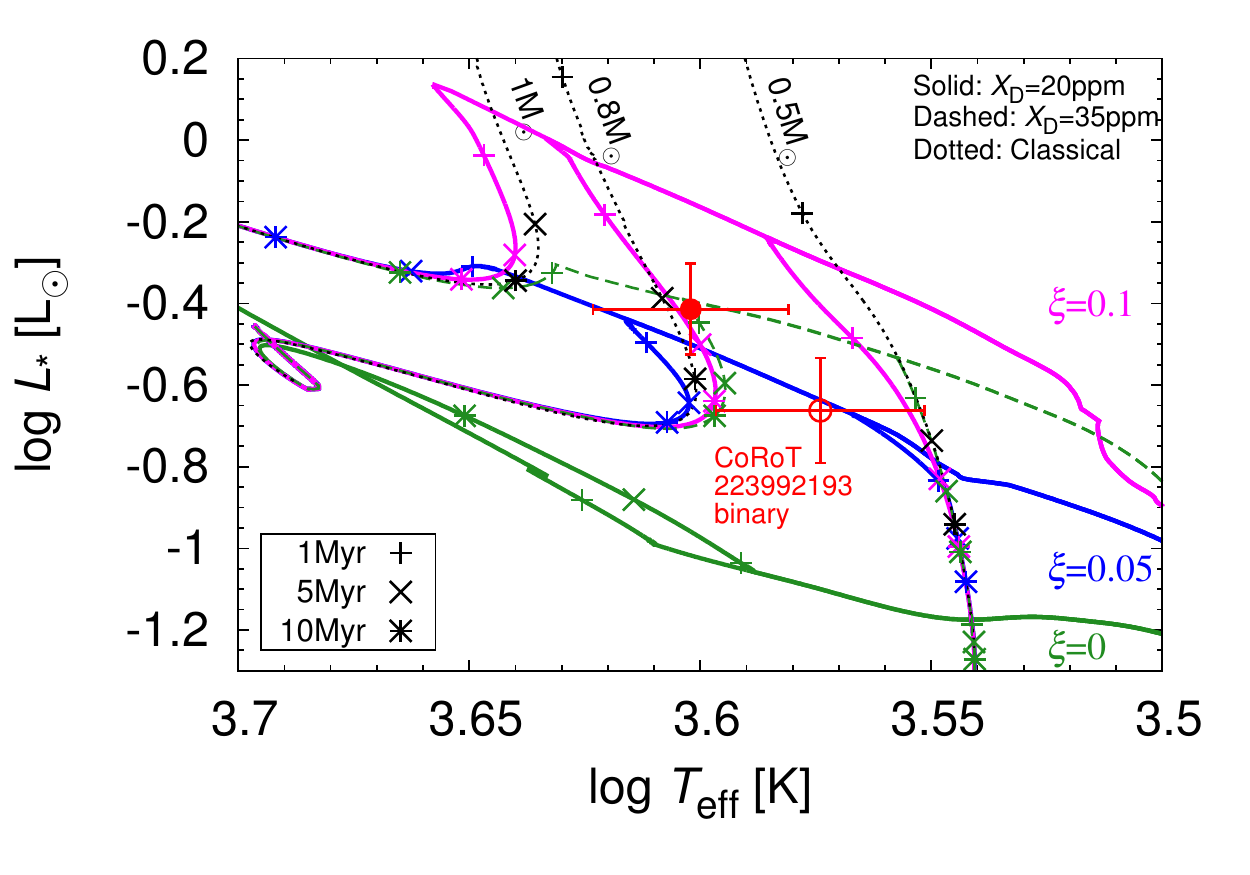}
        \caption{\small{
        Evolutionary tracks whose final masses are 0.5, 0.8, and 1~$\Msun$ for the cases of $\xi=0$ (green lines), 0.05 (blue), and 0.1 (magenta dot-dashed).         The initial radius is $1.5~\Rsun$ and $X\sub{D}=20~\rm{ppm}$ (solid lines) and $35~\rm{ppm}$ (dashed).
        Some are merged to the Hayashi track of the non-accreting evolutionary tracks (black dotted), and others are not. Thus, the mass determination of stars on the Hayashi phase is not affected by the variation in accretion heat.
        As already discussed, the age determination is affected by $\xi$. Pluses, crosses, and asterisks denote 1, 5, and 10 million years, respectively.
        The filled and open circles represent the primary and secondary of the eclipsing binaries CoRoT 223992193, respectively \citep{Stassun+14}. For the cases of $X\sub{D}=20~\rm{ppm}$, their age is estimated to 1 or 5 million years if $\xi=0.05$ or 0.1, respectively.
        }}\label{fig:mass_anomary}
    \end{center}
\end{figure}

Figure~\ref{fig:mass_anomary} shows the constraints on luminosity and effective temperature for both components of the system and compares them to evolutionary tracks for classical (non-accreting) models, and for our models with values of $\xi$ between 0 and 0.1, $X\sub{D}$ of 20 and 35 ppm, and $R\sub{ini}=1.5\,\Rsun$. 
The two stars have masses between $0.6\,\Msun$ and $0.8\,\Msun$ and an age less than 5~Myr.
{{Although the evolutionary tracks are strongly sensitive to the assumed values of $\xi$, 
the isochrones remain mostly parallel to each other, except for extremely low values of $\xi$ with small $X\sub{D}$. This implies that varying $\xi$ essentially results in a shift in age, where stars with a given effective temperature and luminosity are younger for lower values of $\xi$. 
}}

%%%%%%
%%%%%%

Quantitatively, the constraints that we can derive on $\xi$ are very similar to those obtained for young clusters (and with the same caveats). {{With the assumptions of $R\sub{ini}$ and the uniform distribution,}} for $X\sub{D}=20$\,ppm, we obtain that $\xi\ge 0.05,$ with the limiting case corresponding to both stars being close to their birthline. For $X\sub{D}=35$\,ppm, the initial entropy is larger so that no constraint on $\xi$ can be obtained.

%________________________________________________________

\section{Conclusions}\label{sec:conclusion}
The PMS evolution of stars has long been considered a relatively simple theoretical problem governed by the quasi-static contraction of an almost isentropic {{star}}.
We have seen that this evolution phase may in fact be strongly altered by the fact that the stellar envelope must be accreted onto a protostellar seed and that a significant fraction of the energy may be lost in the accretion shock connected to this process. The goal of this paper was to understand what controls the PMS and to derive constraints from the observations. In order to do so, we used simulations with the stellar evolution code MESA that account for a progressive accretion of material onto a forming star. 

We have first shown that beyond classical parameters, such as mass and metallicity, the evolution on the PMS is controlled essentially by {three} parameters: $\xi$, the efficiency at which the gravitational energy of the accreted material is transformed into internal energy of the star, $X\sub{D}$, the mass mixing ratio of deuterium, {and the entropy of the initial stellar seed (or equivalently in the present work its radius $R\sub{ini}$ for a mass of $0.01\,\Msun$)}. The parameter
$\xi=0.5$ corresponds to the classical models and lead to the formation of stars with a large radius {and entropy and an evolution that is essentially independent of the two other parameters, $X\sub{D}$ and $R\sub{ini}$.}
Progressively lower values of $\xi$ yield stars with (much) smaller radii and a richer ensemble of possibilities in terms of their evolution. In particular, because the entropy of the accreted material is then effectively smaller, the abundance of deuterium in that material becomes very important in deciding whether stars will expand significantly (high $X\sub{D}$ values) or will retain a small size all the way to the main sequence (for $X\sub{D}\wig{<}30$\,ppm). We showed that the differences between the results of \citet{BCG09} and \citet{Hosokawa+11} can be accurately reproduced and the results from different choices for $X\sub{D}$.

We compared the evolutionary tracks to the observations of several young clusters.
We {confirmed} that the spread in luminosities in each cluster could be explained without invoking an age spread but by instead assuming that $\xi$ could vary from one star to the next. 
{{
A variation of $R\sub{ini}$ (i.e., stellar entropy at $0.01\,\Msun$) is also needed to explain underluminous, cool stars in Tau and Cha I.
}}
However, the observations indicate that {{most}} stars cannot be too low in entropy when they form.
This implies that within the uniform accretion model, we can rule out as unlikely those scenarios with low $\xi$ and $X\sub{D}$ values. 
Specifically, {{the model with $X\sub{D}=20$\,ppm, $R\sub{ini}=1.5\,\Rsun$, and $\xi=0.1$ sets a useful boundary: stars can have lower entropies only in relatively rare cases. This means that for a majority of stars, stellar evolution cannot differ from the classical evolution tracks beyond the limit set by this limiting model. On the other hand, because of the multi-parameter dependence, we cannot derive an independent constraint on $\xi$. For example, a model with $X\sub{D}=35$\,ppm and $\xi=0$ is equivalent in terms of entropy and luminosities in the H-R diagram to the $X\sub{D}=20$\,ppm and $\xi=0.1$. }}
We found these constraints to be compatible with the observational constraints from the PMS eclipsing binary CoRoT 223992193.
{One caveat is that if a significant number of stars in the clusters are affected by observational errors (such as $L_\star$, membership), our constraints is changed.}

%%%%%
{{Separately, we found that the possibility of reliably measuring deuterium abundances in clusters would allow testing for inefficient accretion (i.e., $\xi<0.5$). If this is the case, we would expect the spread in luminosity to be larger in clusters with a lower deuterium to hydrogen ratio. Present observations indicate that $\sigma$ Ori may have a smaller luminosity spread than other clusters. Although many other factors are to be considered, it is possible that this is due to a higher D/H ratio in that cluster. 
}}
%%%%%

The PMS evolution of stars is not as simple as once thought and merits further investigation. On the observational side, further insight would be gained through the discovery of more very young eclipsing binaries, the determination of deuterium abundances in clusters, and further observations of young accreting stars. On the theoretical side, the main uncertainties in our calculations are due to extremely simplified outer boundary conditions and the ad hoc prescription used to relate the gravitational energy of the accreted material to the internal energy in the star. Three-dimensional radiation hydrodynamic simulations of a collapsing molecular cloud core with sufficient resolution to resolve the central stellar seed are needed.    

A consequence of this more complex PMS evolution is that the stellar interior is not necessarily fully convective during most of this phase. This may have strong implications to understand the chemical composition of stars and connect measurements of stellar compositions to the formation of planets. We will investigate this issue in our next paper.

\begin{acknowledgements}
We express our gratitude to S. Inutsuka, T. Hosokawa, P. Morel, T. Nakamoto, M. Kuzuhara, and M. Ikoma for fruitful discussions and comments.
Bill Paxton and Dean Townsley kindly helped M.K. use the stellar-evolution code MESA.
We appreciate the critical and constructive comments of the {referees}, which helped us to improve this paper.
M.K. is supported by Grant-in-Aid for JSPS Fellows Grant Number 24$\cdot$9296, MEXT of Japan (Grant: 23244027) and Foundation for Promotion of Astronomy.
\end{acknowledgements}

\bibliographystyle{aa}

\begin{thebibliography}{86}
\expandafter\ifx\csname natexlab\endcsname\relax\def\natexlab#1{#1}\fi

\bibitem[{{Allard} {et~al.}(2001){Allard}, {Hauschildt}, {Alexander},
  {Tamanai}, \& {Schweitzer}}]{Allard+01}
{Allard}, F., {Hauschildt}, P.~H., {Alexander}, D.~R., {Tamanai}, A., \&
  {Schweitzer}, A. 2001, \apj, 556, 357

\bibitem[{{Amelin} {et~al.}(2002){Amelin}, {Krot}, {Hutcheon}, \&
  {Ulyanov}}]{Amelin+02}
{Amelin}, Y., {Krot}, A.~N., {Hutcheon}, I.~D., \& {Ulyanov}, A.~A. 2002,
  Science, 297, 1678

\bibitem[{{Angulo} {et~al.}(1999){Angulo}, {Arnould}, {Rayet}, {Descouvemont},
  {Baye}, {Leclercq-Willain}, {Coc}, {Barhoumi}, {Aguer}, {Rolfs}, {Kunz},
  {Hammer}, {Mayer}, {Paradellis}, {Kossionides}, {Chronidou}, {Spyrou},
  {degl'Innocenti}, {Fiorentini}, {Ricci}, {Zavatarelli}, {Providencia},
  {Wolters}, {Soares}, {Grama}, {Rahighi}, {Shotter}, \& {Lamehi
  Rachti}}]{Angulo+99}
{Angulo}, C., {Arnould}, M., {Rayet}, M., {et~al.} 1999, Nuclear Physics A,
  656, 3

\bibitem[{{Asplund} {et~al.}(2009){Asplund}, {Grevesse}, {Sauval}, \&
  {Scott}}]{Asplund+09}
{Asplund}, M., {Grevesse}, N., {Sauval}, A.~J., \& {Scott}, P. 2009, \araa, 47,
  481

\bibitem[{{Bahcall} {et~al.}(2005){Bahcall}, {Basu}, {Pinsonneault}, \&
  {Serenelli}}]{Bahcall+05}
{Bahcall}, J.~N., {Basu}, S., {Pinsonneault}, M., \& {Serenelli}, A.~M. 2005,
  \apj, 618, 1049

\bibitem[{{Baraffe} \& {Chabrier}(2010)}]{BC10}
{Baraffe}, I. \& {Chabrier}, G. 2010, \aap, 521, A44

\bibitem[{{Baraffe} {et~al.}(1998){Baraffe}, {Chabrier}, {Allard}, \&
  {Hauschildt}}]{Baraffe+98}
{Baraffe}, I., {Chabrier}, G., {Allard}, F., \& {Hauschildt}, P.~H. 1998, \aap,
  337, 403

\bibitem[{{Baraffe} {et~al.}(2009){Baraffe}, {Chabrier}, \& {Gallardo}}]{BCG09}
{Baraffe}, I., {Chabrier}, G., \& {Gallardo}, J. 2009, \apjl, 702, L27

\bibitem[{{Baraffe} {et~al.}(2012){Baraffe}, {Vorobyov}, \& {Chabrier}}]{BVC12}
{Baraffe}, I., {Vorobyov}, E., \& {Chabrier}, G. 2012, \apj, 756, 118

\bibitem[{{Basu} \& {Antia}(2004)}]{Basu+Antia04}
{Basu}, S. \& {Antia}, H.~M. 2004, \apjl, 606, L85

\bibitem[{{Burningham} {et~al.}(2005){Burningham}, {Naylor}, {Littlefair}, \&
  {Jeffries}}]{Burningham+05}
{Burningham}, B., {Naylor}, T., {Littlefair}, S.~P., \& {Jeffries}, R.~D. 2005,
  \mnras, 363, 1389

\bibitem[{{Cassisi} {et~al.}(2007){Cassisi}, {Potekhin}, {Pietrinferni},
  {Catelan}, \& {Salaris}}]{Cassisi+07}
{Cassisi}, S., {Potekhin}, A.~Y., {Pietrinferni}, A., {Catelan}, M., \&
  {Salaris}, M. 2007, \apj, 661, 1094

\bibitem[{{Castelli} \& {Kurucz}(2004)}]{Castelli+Kurucz04}
{Castelli}, F. \& {Kurucz}, R.~L. 2004, ArXiv Astrophysics e-prints

\bibitem[{{Chabrier} {et~al.}(2007){Chabrier}, {Gallardo}, \&
  {Baraffe}}]{Chabrier+07}
{Chabrier}, G., {Gallardo}, J., \& {Baraffe}, I. 2007, \aap, 472, L17

\bibitem[{{Chambers}(2010)}]{Chambers10}
{Chambers}, J.~E. 2010, \apj, 724, 92

\bibitem[{{Chandrasekhar}(1967)}]{Chandrasekhar67}
{Chandrasekhar}, S. 1967, {An introduction to the study of stellar structure}

\bibitem[{Cox \& Giuli(1968)}]{Cox+Giuli68}
Cox, J. \& Giuli, R. 1968, Gordon and Breach, New York, 401

\bibitem[{{Da Rio} {et~al.}(2010){Da Rio}, {Robberto}, {Soderblom}, {Panagia},
  {Hillenbrand}, {Palla}, \& {Stassun}}]{Da-Rio+10}
{Da Rio}, N., {Robberto}, M., {Soderblom}, D.~R., {et~al.} 2010, \apj, 722,
  1092

\bibitem[{{Dunham} \& {Vorobyov}(2012)}]{Dunham+Vorobyov12}
{Dunham}, M.~M. \& {Vorobyov}, E.~I. 2012, \apj, 747, 52

\bibitem[{{Ferguson} {et~al.}(2005){Ferguson}, {Alexander}, {Allard}, {Barman},
  {Bodnarik}, {Hauschildt}, {Heffner-Wong}, \& {Tamanai}}]{Ferguson+05}
{Ferguson}, J.~W., {Alexander}, D.~R., {Allard}, F., {et~al.} 2005, \apj, 623,
  585

\bibitem[{{Gatti} {et~al.}(2008){Gatti}, {Natta}, {Randich}, {Testi}, \&
  {Sacco}}]{Gatti+08}
{Gatti}, T., {Natta}, A., {Randich}, S., {Testi}, L., \& {Sacco}, G. 2008,
  \aap, 481, 423

\bibitem[{{Gatti} {et~al.}(2006){Gatti}, {Testi}, {Natta}, {Randich}, \&
  {Muzerolle}}]{Gatti+06}
{Gatti}, T., {Testi}, L., {Natta}, A., {Randich}, S., \& {Muzerolle}, J. 2006,
  \aap, 460, 547

\bibitem[{{Geroux} {et~al.}(2016){Geroux}, {Baraffe}, {Viallet}, {Goffrey},
  {Pratt}, {Constantino}, {Folini}, {Popov}, \& {Walder}}]{Geroux+16}
{Geroux}, C., {Baraffe}, I., {Viallet}, M., {et~al.} 2016, \aap, 588, A85

\bibitem[{{Gillen} {et~al.}(2014){Gillen}, {Aigrain}, {McQuillan}, {Bouvier},
  {Hodgkin}, {Alencar}, {Terquem}, {Southworth}, {Gibson}, {Cody}, {Lendl},
  {Morales-Calder{\'o}n}, {Favata}, {Stauffer}, \& {Micela}}]{Gillen+14}
{Gillen}, E., {Aigrain}, S., {McQuillan}, A., {et~al.} 2014, \aap, 562, A50

\bibitem[{{Grevesse} \& {Sauval}(1998)}]{GS98}
{Grevesse}, N. \& {Sauval}, A.~J. 1998, \ssr, 85, 161

\bibitem[{{Guillot}(1999)}]{Guillot99}
{Guillot}, T. 1999, \planss, 47, 1183

\bibitem[{{Guillot} {et~al.}(2014){Guillot}, {Ida}, \& {Ormel}}]{Guillot+14}
{Guillot}, T., {Ida}, S., \& {Ormel}, C.~W. 2014, \aap, 572, A72

\bibitem[{{Hartmann}(2001)}]{Hartmann01}
{Hartmann}, L. 2001, \aj, 121, 1030

\bibitem[{{Hartmann} {et~al.}(1997){Hartmann}, {Cassen}, \&
  {Kenyon}}]{Hartmann+97}
{Hartmann}, L., {Cassen}, P., \& {Kenyon}, S.~J. 1997, \apj, 475, 770

\bibitem[{{Hauschildt} {et~al.}(1999{\natexlab{a}}){Hauschildt}, {Allard}, \&
  {Baron}}]{Hauschildt+99a}
{Hauschildt}, P.~H., {Allard}, F., \& {Baron}, E. 1999{\natexlab{a}}, \apj,
  512, 377

\bibitem[{{Hauschildt} {et~al.}(1999{\natexlab{b}}){Hauschildt}, {Allard},
  {Ferguson}, {Baron}, \& {Alexander}}]{Hauschildt+99b}
{Hauschildt}, P.~H., {Allard}, F., {Ferguson}, J., {Baron}, E., \& {Alexander},
  D.~R. 1999{\natexlab{b}}, \apj, 525, 871

\bibitem[{{Hayashi}(1961)}]{Hayashi61}
{Hayashi}, C. 1961, \pasj, 13, 450

\bibitem[{{Heber} {et~al.}(2008){Heber}, {Baur}, {Bochsler}, {Burnett},
  {Reisenfeld}, {Wieler}, \& {Wiens}}]{Heber+08}
{Heber}, V.~S., {Baur}, H., {Bochsler}, P., {et~al.} 2008, in Lunar and
  Planetary Inst. Technical Report, Vol.~39, Lunar and Planetary Science
  Conference, 1779

\bibitem[{{H{\'e}brard} {et~al.}(2005){H{\'e}brard}, {Tripp}, {Chayer},
  {Friedman}, {Dupuis}, {Sonnentrucker}, {Williger}, \& {Moos}}]{Hebrard+05}
{H{\'e}brard}, G., {Tripp}, T.~M., {Chayer}, P., {et~al.} 2005, \apj, 635, 1136

\bibitem[{{Henyey} {et~al.}(1965){Henyey}, {Vardya}, \&
  {Bodenheimer}}]{Henyey+65}
{Henyey}, L., {Vardya}, M.~S., \& {Bodenheimer}, P. 1965, \apj, 142, 841

\bibitem[{{Herwig}(2000)}]{Herwig00}
{Herwig}, F. 2000, \aap, 360, 952

\bibitem[{{Hillenbrand}(2009)}]{Hillenbrand09}
{Hillenbrand}, L.~A. 2009, in IAU Symposium, Vol. 258, IAU Symposium, ed. E.~E.
  {Mamajek}, D.~R. {Soderblom}, \& R.~F.~G. {Wyse}, 81--94

\bibitem[{{Hosokawa} {et~al.}(2011){Hosokawa}, {Offner}, \&
  {Krumholz}}]{Hosokawa+11}
{Hosokawa}, T., {Offner}, S.~S.~R., \& {Krumholz}, M.~R. 2011, \apj, 738, 140

\bibitem[{{Hosokawa} \& {Omukai}(2009)}]{Hosokawa+Omukai09}
{Hosokawa}, T. \& {Omukai}, K. 2009, \apj, 691, 823

\bibitem[{{Hueso} \& {Guillot}(2005)}]{Hueso+Guillot05}
{Hueso}, R. \& {Guillot}, T. 2005, \aap, 442, 703

\bibitem[{{Inutsuka}(2012)}]{Inutsuka12}
{Inutsuka}, S.-i. 2012, Progress of Theoretical and Experimental Physics, 2012,
  010000

\bibitem[{{Inutsuka} {et~al.}(2015){Inutsuka}, {Inoue}, {Iwasaki}, \&
  {Hosokawa}}]{Inutsuka+15}
{Inutsuka}, S.-i., {Inoue}, T., {Iwasaki}, K., \& {Hosokawa}, T. 2015, \aap,
  580, A49

\bibitem[{{Inutsuka} {et~al.}(2010){Inutsuka}, {Machida}, \&
  {Matsumoto}}]{Inutsuka+10}
{Inutsuka}, S.-i., {Machida}, M.~N., \& {Matsumoto}, T. 2010, \apjl, 718, L58

\bibitem[{{Jeffries}(2007)}]{Jeffries07}
{Jeffries}, R.~D. 2007, \mnras, 381, 1169

\bibitem[{{Jeffries}(2012)}]{Jeffries12}
{Jeffries}, R.~D. 2012, {Are There Age Spreads in Star Forming Regions?}, ed.
  A.~{Moitinho} \& J.~{Alves}, 163

\bibitem[{{Kenyon} \& {Hartmann}(1995)}]{Kenyon+Hartmann95}
{Kenyon}, S.~J. \& {Hartmann}, L. 1995, \apjs, 101, 117

\bibitem[{{Larson}(1969)}]{Larson69}
{Larson}, R.~B. 1969, \mnras, 145, 271

\bibitem[{{Lellouch} {et~al.}(2001){Lellouch}, {B{\'e}zard}, {Fouchet},
  {Feuchtgruber}, {Encrenaz}, \& {de Graauw}}]{Lellouch+01}
{Lellouch}, E., {B{\'e}zard}, B., {Fouchet}, T., {et~al.} 2001, \aap, 370, 610

\bibitem[{{Linsky} {et~al.}(2006){Linsky}, {Draine}, {Moos}, {Jenkins}, {Wood},
  {Oliveira}, {Blair}, {Friedman}, {Gry}, {Knauth}, {Kruk}, {Lacour}, {Lehner},
  {Redfield}, {Shull}, {Sonneborn}, \& {Williger}}]{Linsky+06}
{Linsky}, J.~L., {Draine}, B.~T., {Moos}, H.~W., {et~al.} 2006, \apj, 647, 1106

\bibitem[{{Mahaffy} {et~al.}(1998){Mahaffy}, {Donahue}, {Atreya}, {Owen}, \&
  {Niemann}}]{Mahaffy+98}
{Mahaffy}, P.~R., {Donahue}, T.~M., {Atreya}, S.~K., {Owen}, T.~C., \&
  {Niemann}, H.~B. 1998, \ssr, 84, 251

\bibitem[{{Martin} {et~al.}(2012){Martin}, {Lubow}, {Livio}, \&
  {Pringle}}]{Martin+12}
{Martin}, R.~G., {Lubow}, S.~H., {Livio}, M., \& {Pringle}, J.~E. 2012, \mnras,
  423, 2718

\bibitem[{{Masunaga} \& {Inutsuka}(2000)}]{Masunaga+Inutsuka00}
{Masunaga}, H. \& {Inutsuka}, S.-i. 2000, \apj, 531, 350

\bibitem[{{Mel{\'e}ndez} {et~al.}(2009){Mel{\'e}ndez}, {Asplund}, {Gustafsson},
  \& {Yong}}]{Melendez+09}
{Mel{\'e}ndez}, J., {Asplund}, M., {Gustafsson}, B., \& {Yong}, D. 2009, \apjl,
  704, L66

\bibitem[{{Mercer-Smith} {et~al.}(1984){Mercer-Smith}, {Cameron}, \&
  {Epstein}}]{Mercer-Smith+84}
{Mercer-Smith}, J.~A., {Cameron}, A.~G.~W., \& {Epstein}, R.~I. 1984, \apj,
  279, 363

\bibitem[{{Muzerolle} {et~al.}(2005){Muzerolle}, {Luhman}, {Brice{\~n}o},
  {Hartmann}, \& {Calvet}}]{Muzerolle+05}
{Muzerolle}, J., {Luhman}, K.~L., {Brice{\~n}o}, C., {Hartmann}, L., \&
  {Calvet}, N. 2005, \apj, 625, 906

\bibitem[{Nelder \& Mead(1965)}]{Nelder+Mead65}
Nelder, J.~A. \& Mead, R. 1965, The computer journal, 7, 308

\bibitem[{{Palla} \& {Stahler}(1992)}]{Palla+Stahler92}
{Palla}, F. \& {Stahler}, S.~W. 1992, \apj, 392, 667

\bibitem[{{Palla} \& {Stahler}(2000)}]{Palla+Stahler00}
{Palla}, F. \& {Stahler}, S.~W. 2000, \apj, 540, 255

\bibitem[{{Paxton} {et~al.}(2011){Paxton}, {Bildsten}, {Dotter}, {Herwig},
  {Lesaffre}, \& {Timmes}}]{Paxton+11}
{Paxton}, B., {Bildsten}, L., {Dotter}, A., {et~al.} 2011, \apjs, 192, 3

\bibitem[{{Paxton} {et~al.}(2013){Paxton}, {Cantiello}, {Arras}, {Bildsten},
  {Brown}, {Dotter}, {Mankovich}, {Montgomery}, {Stello}, {Timmes}, \&
  {Townsend}}]{Paxton+13}
{Paxton}, B., {Cantiello}, M., {Arras}, P., {et~al.} 2013, \apjs, 208, 4

\bibitem[{{Paxton} {et~al.}(2015){Paxton}, {Marchant}, {Schwab}, {Bauer},
  {Bildsten}, {Cantiello}, {Dessart}, {Farmer}, {Hu}, {Langer}, {Townsend},
  {Townsley}, \& {Timmes}}]{Paxton+15}
{Paxton}, B., {Marchant}, P., {Schwab}, J., {et~al.} 2015, \apjs, 220, 15

\bibitem[{{Prantzos}(2007)}]{Prantzos+07}
{Prantzos}, N. 2007, \ssr, 130, 27

\bibitem[{{Ram{\'{\i}}rez} {et~al.}(2009){Ram{\'{\i}}rez}, {Mel{\'e}ndez}, \&
  {Asplund}}]{Ramirez+09}
{Ram{\'{\i}}rez}, I., {Mel{\'e}ndez}, J., \& {Asplund}, M. 2009, \aap, 508, L17

\bibitem[{{Ram{\'{\i}}rez} {et~al.}(2011){Ram{\'{\i}}rez}, {Mel{\'e}ndez},
  {Cornejo}, {Roederer}, \& {Fish}}]{Ramirez+11}
{Ram{\'{\i}}rez}, I., {Mel{\'e}ndez}, J., {Cornejo}, D., {Roederer}, I.~U., \&
  {Fish}, J.~R. 2011, \apj, 740, 76

\bibitem[{{Reggiani} {et~al.}(2011){Reggiani}, {Robberto}, {Da Rio}, {Meyer},
  {Soderblom}, \& {Ricci}}]{Reggiani+11}
{Reggiani}, M., {Robberto}, M., {Da Rio}, N., {et~al.} 2011, \aap, 534, A83

\bibitem[{{Rogers} \& {Nayfonov}(2002)}]{Rogers+Nayfonov02}
{Rogers}, F.~J. \& {Nayfonov}, A. 2002, \apj, 576, 1064

\bibitem[{{Saumon} {et~al.}(1995){Saumon}, {Chabrier}, \& {van
  Horn}}]{Saumon+95}
{Saumon}, D., {Chabrier}, G., \& {van Horn}, H.~M. 1995, \apjs, 99, 713

\bibitem[{{Seaton}(2005)}]{Seaton05}
{Seaton}, M.~J. 2005, \mnras, 362, L1

\bibitem[{{Shu}(1977)}]{Shu77}
{Shu}, F.~H. 1977, \apj, 214, 488

\bibitem[{{Soderblom} {et~al.}(2014){Soderblom}, {Hillenbrand}, {Jeffries},
  {Mamajek}, \& {Naylor}}]{Soderblom+14}
{Soderblom}, D.~R., {Hillenbrand}, L.~A., {Jeffries}, R.~D., {Mamajek}, E.~E.,
  \& {Naylor}, T. 2014, Protostars and Planets VI, 219

\bibitem[{{Stahler}(1983)}]{Stahler+83}
{Stahler}, S.~W. 1983, \apj, 274, 822

\bibitem[{{Stahler}(1988)}]{Stahler88}
{Stahler}, S.~W. 1988, \apj, 332, 804

\bibitem[{{Stahler} \& {Palla}(2005)}]{Stahler+Palla05}
{Stahler}, S.~W. \& {Palla}, F. 2005, {The Formation of Stars}

\bibitem[{{Stahler} {et~al.}(1986){Stahler}, {Palla}, \&
  {Salpeter}}]{Stahler+86}
{Stahler}, S.~W., {Palla}, F., \& {Salpeter}, E.~E. 1986, \apj, 302, 590

\bibitem[{{Stahler} {et~al.}(1980){Stahler}, {Shu}, \& {Taam}}]{SST80I}
{Stahler}, S.~W., {Shu}, F.~H., \& {Taam}, R.~E. 1980, \apj, 241, 637

\bibitem[{{Stamatellos} {et~al.}(2007){Stamatellos}, {Whitworth}, {Bisbas}, \&
  {Goodwin}}]{Stamatellos+07}
{Stamatellos}, D., {Whitworth}, A.~P., {Bisbas}, T., \& {Goodwin}, S. 2007,
  \aap, 475, 37

\bibitem[{{Stassun} {et~al.}(2014){Stassun}, {Feiden}, \&
  {Torres}}]{Stassun+14}
{Stassun}, K.~G., {Feiden}, G.~A., \& {Torres}, G. 2014, \nar, 60, 1

\bibitem[{{Steigman}(2006)}]{Steigman06}
{Steigman}, G. 2006, International Journal of Modern Physics E, 15, 1

\bibitem[{{Sugimoto} \& {Nomoto}(1975)}]{Sugimoto+Nomoto75}
{Sugimoto}, D. \& {Nomoto}, K. 1975, \pasj, 27, 197

\bibitem[{{Tomida} {et~al.}(2013){Tomida}, {Tomisaka}, {Matsumoto}, {Hori},
  {Okuzumi}, {Machida}, \& {Saigo}}]{Tomida+13}
{Tomida}, K., {Tomisaka}, K., {Matsumoto}, T., {et~al.} 2013, \apj, 763, 6

\bibitem[{{Townsley} \& {Bildsten}(2004)}]{Townsley+Bildsten04}
{Townsley}, D.~M. \& {Bildsten}, L. 2004, \apj, 600, 390

\bibitem[{{Vaytet} {et~al.}(2013){Vaytet}, {Chabrier}, {Audit}, {Commer{\c
  c}on}, {Masson}, {Ferguson}, \& {Delahaye}}]{Vaytet+13}
{Vaytet}, N., {Chabrier}, G., {Audit}, E., {et~al.} 2013, \aap, 557, A90

\bibitem[{{Vidal-Madjar} \& {Gry}(1984)}]{Vidal-Madjar+Gry84}
{Vidal-Madjar}, A. \& {Gry}, C. 1984, \aap, 138, 285

\bibitem[{{Vorobyov} \& {Basu}(2005)}]{VB05}
{Vorobyov}, E.~I. \& {Basu}, S. 2005, \apjl, 633, L137

\bibitem[{{Vorobyov} \& {Basu}(2010)}]{Vorobyov+Basu10}
{Vorobyov}, E.~I. \& {Basu}, S. 2010, \apj, 719, 1896

\bibitem[{{Winkler} \& {Newman}(1980)}]{Winkler+Newman80}
{Winkler}, K.-H.~A. \& {Newman}, M.~J. 1980, \apj, 236, 201

\end{thebibliography}

\begin{appendix} %First online appendix

%--------------
\section{$\chi^2$ test for the input parameters} \label{app:param}

In this study we chose the input parameters that reproduce the observed solar quantities.
In addition to the radius and luminosity, the internal structure and surface composition are constrained by helioseismic and spectroscopic analyses \citep[][and references therein]{Bahcall+05, Basu+Antia04,GS98}.
Although the solar metallicity remains a matter of debate\footnote{
Using three-dimensional atmosphere model, the metallicity in the solar atmosphere is estimated to be $\sim 0.0134$, which is reduced from the classical value ($\sim 0.02$) by about 70\% \citep[see][]{Asplund+09}.}
it is beyond the scope of this paper.

We performed a $\chi^2$ test to find the best initial settings using the ``Nelder-Mead simplex algorithm'' \citep[][]{Nelder+Mead65}. 
The input parameters are the initial composition ($X_{\mathrm{ini}}$, $Y_{\mathrm{ini}}$, and $Z_{\mathrm{ini}}$; see Sect.~\ref{sec:chem}), the mixing-length parameter, $\alpha_{\mathrm{MLT}}$, and the overshoot mixing parameter, $f_\mathrm{ov}$ \citep[][see Sect.~\ref{sec:code}]{Herwig00, Paxton+11}.

We calculated the evolution of $1~\Msun$ stars from PMS phase with changing these parameters to minimize the $\chi^2$ value at the solar age, which is assumed to be 4.567~Gyr \citep{Amelin+02}.
The results of the $\chi^2$ test listed in Table~\ref{Table:solar} are used in this paper.

%__________________________________________________ One column table
   \begin{table}[!htb]
      \caption[]{Results of the $\chi^2$ test to reproduce solar values.}
         \label{Table:solar} \small
         \begin{tabular}{p{1.5cm}lp{2.6cm}}
            \hline
            \hline
%%%%%%
            \noalign{\smallskip}
            & \multicolumn{2}{l}{Converged input parameters} \\
            \noalign{\smallskip}
            \hline
            \noalign{\smallskip}
%%%%%%
            $X_{\mathrm{ini}}$                  & 0.7004553948 & \\
            $Y_{\mathrm{ini}}$                  & 0.2794811789 & \\
            $Z_{\mathrm{ini}}$                  & 0.0200634263 & \\
            $\alpha\sub{MLT}$               & 1.9050629261 & \\
            $f_{\mathrm{ov}}$                   & 0.0119197042 & \\
%%%%%%
            \noalign{\smallskip}
            \hline
            \noalign{\smallskip}
             & Target values & Converged values at Solar age\\
            \noalign{\smallskip}
            \hline
            \noalign{\smallskip}
            $(Z/X)_{\mathrm{surf}}$             & $0.02313\tablefootmark{a}$ & 0.02534371   \\
            $Y_{\mathrm{surf}}$                 & $0.2485\pm0.0035\tablefootmark{b}$ & 0.25338885 \\
            $R_{\mathrm{CZ}}/\Rsun$     & $0.713\pm0.001\tablefootmark{c}$ & 0.71275778   \\
            $R_{\star}$                 & $6.9598 \times 10^{10}~{\mathrm{cm}}\tablefootmark{c}$  & $1.00000282~\Rsun$  \\
            $L_{\star}$                 & $3.8418 \times 10^{33}~{\mathrm{erg/s}}\tablefootmark{c}$ & $1.00000114~\Lsun$  \\
            $T_{\mathrm{eff}}~[{\textrm{K}} ]$                  & $5776.13157\tablefootmark{d}$ & $5776.02320$   \\
%%%%%%
            \noalign{\smallskip}
            \hline
            \noalign{\smallskip}
            \multicolumn{3}{p{8.3cm}}{\textbf{Notes.} $^{(a)}$ \citet{GS98}. $^{(b)}$ \citet{Basu+Antia04}. $^{(c)}$ \citet{Bahcall+05}. $^{(d)}$ Calculated with $L_\star$, $R_\star$ and Stefan-Boltzmann law.}
         \end{tabular} 
        \normalsize
            \end{table}
%--------------

%--------------
\section{Evolution and underlying physics in the cold accretion case} \label{sec:overview}

{{In this Appendix, we explain the basic behavior and its underlying physics in the case of cold accretion shown in Sect.\,\ref{sec:typical}.
The evolution can be split into five phases as shown in Fig.\,\ref{fig:1d-5t-R}. We explain each phase below.
}}

\

\begin{figure}[!tb]
  \begin{center}
    \includegraphics[width=\hsize,keepaspectratio,clip]{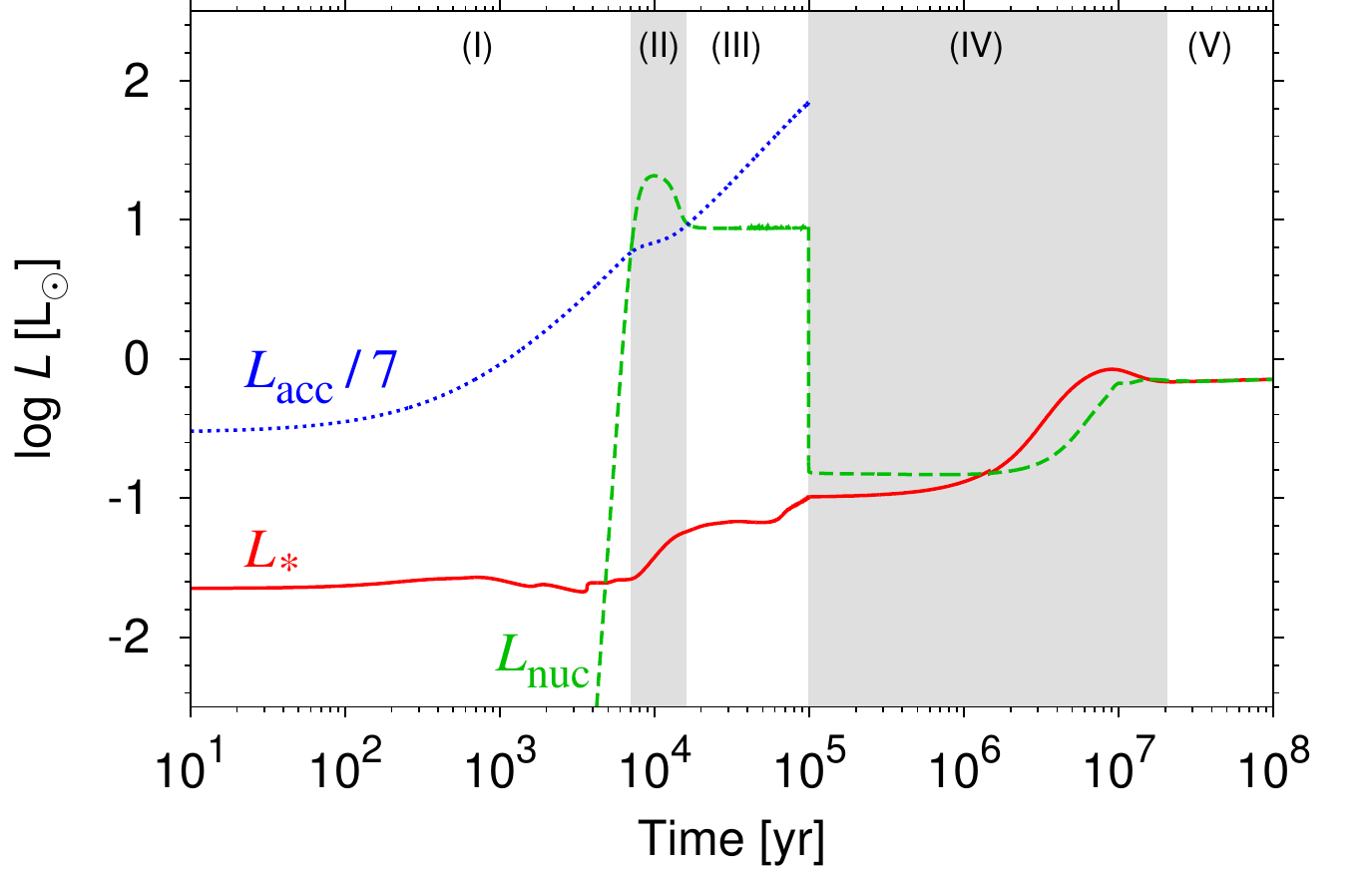}
        \caption{\small{Evolution of the intrinsic luminosity ($L_\star$, solid line), the energy production rate by thermonuclear reaction ($L\sub{nuc}$, dashed) and the seventh part of the gravitational energy of the accreting material ($\frac{1}{7}L\sub{acc}$, dotted). The shaded regions and labels from (I) to (V) are the same as in Fig.~\ref{fig:1d-5t-R}.
        During phase (II), $L\sub{nuc}$ exceeds $\frac{1}{7}L\sub{acc}$ and the star expands.
        The flat $L\sub{nuc}$ during phase (III) results from the steady and instantaneous fusion of the newly accreted deuterium.}}\label{fig:1d-5-t-L}
    \end{center}
\end{figure}

%%%
\textit{I. The adiabatic contraction phase:}\\
In this phase, the star shrinks while increasing its mass. The radius evolution can be fitted by
\begin{equation}
R_\star \propto M_\star^{-1/3}~. \label{eq:R-M}
\end{equation}
Indeed, we show in Appendix~\ref{app:anal} that this relation is verified for a perfect gas star that accretes {{gas with the same entropy.}}
In the current settings, the accretion is intense enough that the Kelvin-Helmholtz (K-H) timescale, $\tau\sub{KH} \equiv |E\sub{tot}|/L_\star$, is much longer than the accretion timescale, $\tau\sub{acc} \equiv M_\star/\dot{M}$, where $E\sub{tot}$ is the total energy of the star and $L_\star$ is the stellar intrinsic luminosity \footnote{
The total luminosity of protostars is the sum of the intrinsic luminosity, $L_\star$, and the radiation from the accretion shock front (i.e., $L\sub{acc} - L\sub{add}$).
}.
This is equivalent to $L\sub{acc} \gg L_\star$ as shown in Fig.~\ref{fig:1d-5-t-L}.
The accretion luminosity is defined as 
{{
\begin{eqnarray} \label{eq:Lacc}
L\sub{acc} &\equiv & GM_\star\dot{M}/R_\star\,,
\end{eqnarray}
which is for example $31.3~{\Lsun}$ in the case of $M_\star=0.1\,\Msun$, $R_\star=1\,\Rsun$, and $\dot{M}=10^{-5}~\Msun/\mathrm{yr}$.
}}

%%%%%%
%%%%%%

\

%--------------
\textit{II. The deuterium-burning phase:}\\
After the central temperature exceeds $\sim 10^6~\textrm{K}$, deuterium fusion affects the evolution (see Sect.~\ref{sec:nuc}). In the current settings, it happens at $t = 7 \times 10^3~{\textrm{yr}}$ and $M_\star \simeq 0.08~\Msun$.
As described in Sect.~\ref{sec:nuc}, the energy production rate of deuterium burning has a strong temperature dependence, that is, $\varepsilon\sub{nuc} \propto (T/10^6~{\textrm{K}})^{11.8}$. 
This strong temperature sensitivity 
is responsible for a rapid expansion of the star through the so-called ``thermostat effect'' \citep[e.g.,][]{Stahler88}. 

Indeed, after the ignition of deuterium fusion, the central temperature is maintained constant at about $10^6~\textrm{K}$ by the fact that any temperature increase would result in an expansion of the deuterium burning region, and its adiabatic cooling and  a temperature decrease would be balanced by adiabatic heating, respectively.
With the approximate relations for a perfect gas star in Appendix~\ref{app:anal}, $T\sub{c}\propto P\sub{c}/\rho\sub{c}\propto M_\star/R_\star$ and therefore a constant central temperature implies that a mass increase results in an expansion of the star.

However, the central temperature is not exactly constant so we must test that
the energy produced by deuterium burning is sufficient to cause the expansion. 
In Appendix~\ref{app:anal}, we derive Eq.~\eqref{eq:KHMacc}, which shows how the rate of change of the stellar radius of a fully convective perfect gas star depends on the mass accretion rate, intrinsic luminosity, and nuclear energy production rate. 
Our derivation follows \citet{Hartmann+97}.
By neglecting the intrinsic luminosity (since the Kelvin-Helmholtz timescale is long compared to relevant timescales in this phase),
%%%
\begin{equation}
\frac{\dot{R}}{R_\star} = -\frac{1}{3}\frac{\dot{M}}{M_\star}+\frac{7}{3}\frac{R_\star L\sub{nuc}}{GM_\star^2}\,.\label{eq:lnuccrit}
\end{equation}
%%%
Therefore, the condition for the expansion is given by
%%%
\begin{equation}
L\sub{nuc} > \frac{1}{7}L\sub{acc}\,. \label{eq:lnuccrit2}
\end{equation}
%%%
In the current phase, since this condition is satisfied owing to the vigorous burning of the pre-existent deuterium, $\dot{R}>0$. 

\

%--------------
\textit{III. A second contraction phase:}\\
After $\sim 1.5 \times 10^4~{\textrm{yr}}$, pre-existing deuterium has been burned up and deuterium fusion relies on the accretion of  fresh deuterium. In this phase, the star contracts again.
This is because the mass accretion rate of freshly accreted deuterium, $\dot{M} X\sub{D}$, is not sufficient to compensate for the compression by accretion.
The maximum energy production rate of deuterium burning, $L\sub{D,\,max}$, is calculated by assuming that it is instantaneously burned, 
\begin{equation}
L\sub{D,max}=8.65~L_\odot \brafracket{\dot{M}}{10^{-5}~\Msun/{\textrm{yr}}}\brafracket{X\sub{D}}{2.0\times10^{-5}}~.
\end{equation}
This may be compared to the accretion luminosity $L\sub{acc}$ defined in Eq.~\eqref{eq:Lacc}, but for a mass that is now $M_\star \sim 0.5~\Msun$. We thus obtain $L\sub{D,max}/L\sub{acc}\sim 0.05$ for standard values of the parameters. According to Eq.~\eqref{eq:lnuccrit2}, this shows that the burning of accreted deuterium cannot prevent the adiabatic contraction, independent of the accretion rate. In this phase, thermonuclear energy production is dominated by deuterium burning, implying that $L\sub{nuc}\simeq L\sub{D,max}$ until the accretion ceases at $10^5~{\textrm{yr}}$ (see Fig.~\ref{fig:1d-5-t-L}).

\

%--------------
\textit{IV. The swelling phase:}\\
After the accretion ceases, the radius remains nearly constant for several million years.
This timescale is then determined by what happens in the deep interior and is shorter than the photospheric K-H timescale of $\sim 10^7$--$10^8$~years. 
This is because the luminosity in the deep interior is much larger than that at the surface because of the absorption of energy in the stellar interior.
Thus, the internal thermal timescale, which is $\sim GM_r^2/(2rl)$, is down to a few million years in the deep interior, where $r$ and $l$ are the radius and luminosity.
For most of this phase, the dominant source of the luminosity is the energy release in each mass shell, i.e., $T\partial S/\partial t$ in Eq.~(\ref{eq:dLdM}).

\begin{figure}[!tb]
  \begin{center}
    \includegraphics[width=\hsize,keepaspectratio,clip]{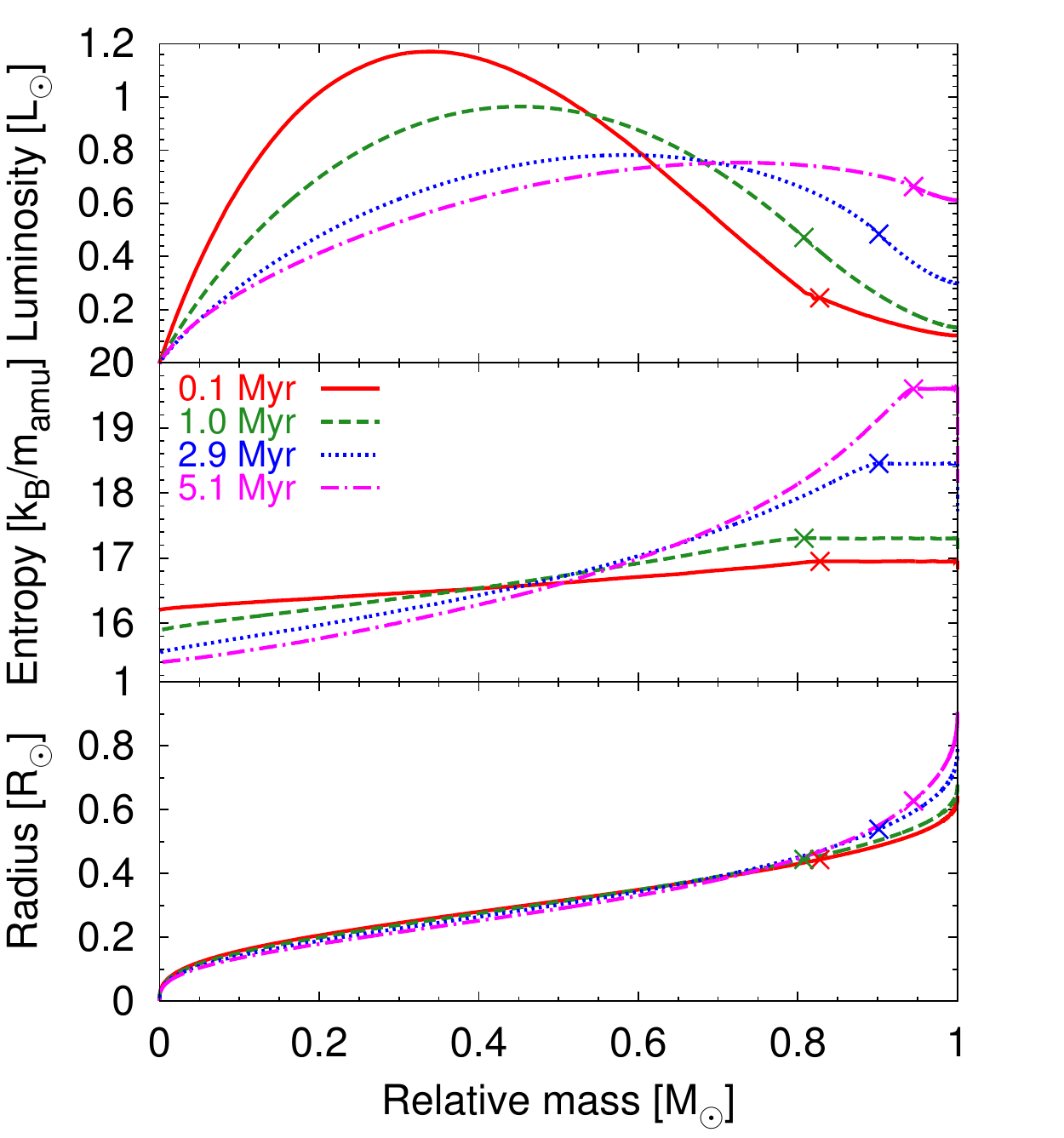}
        \caption{\small{
        Profiles of luminosity (\textit{top panel}), entropy {{(\textit{middle}), and radius}} (\textit{bottom}) in the stellar interior at 0.1 (red solid line; just after the accretion is completed), 1.0 (green dashed), 2.9 (blue dotted), and 5.1 (magenta dot-dashed) million years. As the luminosity propagates toward the surface, the entropy is redistributed. Crosses represent the base of the surface convective zone; $k\sub{B}$ and $m\sub{amu}$ are the Boltzmann constant and the atomic mass unit, respectively.
                }}\label{fig:m-LS}
    \end{center}
\end{figure}

After several million years, the star expands. 
A ``luminosity wave'' \citep[][]{Stahler+86} gradually propagates from the interior to the atmosphere and accompanies an increase in temperature and decrease in opacity of the outer layers.
The redistribution of entropy in the star due to the luminosity wave changes the internal structure (e.g., the polytropic index $n$) and then causes the stellar expansion \citep[][]{Hosokawa+Omukai09}.
These profiles are shown in Fig.~\ref{fig:m-LS}.

%%%%%%
%%%%%%

\

%--------------
\textit{V. The main sequence:}\\
The star then slightly shrinks from the K-H contraction to enter the main sequence (MS).
In this phase, the intrinsic luminosity is almost entirely due to hydrogen burning. The energy equation becomes almost time independent, which means that stars evolve on a much longer timescale. This slow evolution is caused by the change in chemical composition of the central regions due to nuclear energy production. The MS lasts until, in the central regions, hydrogen is exhausted.

%___________________________________________
\section{Analytical relations} \label{app:anal}
Here we derive the mass-radius relationship of Eq.~(\ref{eq:R-M}) in two ways \footnote{In addition, another derivation is possible using a polytropic analysis \citep{Chandrasekhar67}.}.
First, we use the characteristic density and pressure and the entropy following \citet{Stahler88}.
The characteristic density and pressure of the star in the hydrostatic equilibrium are given by
\begin{align}
\tilde{\rho} &\propto M_\star/R_\star^3~, &\tilde{P}&\propto M_\star^2/R_\star^4 \label{eq:rhoP}~.&&
\end{align}
The entropy is given by {{$S=c_V\ln{\left( \frac{P}{\rho^{\gamma\sub{ad}}}\right) }+S_0$, }}
where $c_V$ is the specific heat at constant volume, $\gamma\sub{ad} =c_P/c_V$ the specific heat ratio, and $S_0$ is the constant. Substituting Eq.~(\ref{eq:rhoP}) into this equation, we obtain
\begin{equation}
R_\star=M_\star^{-\frac{2-\gamma\sub{ad}}{3\gamma\sub{ad}-4}} \exp \left( \frac{1}{3\gamma\sub{ad}-4} \frac{S-S_0}{c_V}  \right)~.
\end{equation}
In the case of the monatomic ideal gas ($\gamma\sub{ad}=5/3$), we obtain
\begin{equation} \label{eq:R-M2}
R_\star= M_\star^{-1/3} \exp\left[ \frac{2}{3}\frac{\mu}{\mathcal{R}} (S-S_0)   \right]~,
\end{equation}
where $\mu$ is the mean molecular weight and $\mathcal{R}$ is the gas constant.
This equation shows that the radius is determined only by entropy and mass.
In isentropic stars, $R_\star\propto M_\star^{-1/3}$ (Eq.~\ref{eq:R-M}).

%%%
Secondly, Eq.~(\ref{eq:R-M}) can be derived from the energy equation, which also gives Eq.~(\ref{eq:lnuccrit}) following \citet{Hartmann+97}.
The total energy of a star is given by the sum of the internal and gravitational energy, $E\sub{tot}=E\sub{int}+E\sub{g}$\,.
The energy conservation is expressed as follows:
\begin{equation} \label{eq:dEdt}
\frac{{\mathrm{d}}E\sub{tot}}{\mathrm{d}t} =-L_\star +L\sub{nuc} -\frac{GM_\star\dot{M}}{R_\star}+L\sub{add}\,.
\end{equation}
The right-hand side (RHS) shows the radiative cooling, the energy production by thermonuclear reactions, and gravitational and internal energies of accreting materials. From Eq.~(\ref{eq:Ladd}) here we assume $L\sub{add}=\xi GM_\star\dot{M}/R_\star$\,. 
From the virial theorem,
%%%
\begin{equation}
E\sub{tot}=(4-3\gamma\sub{ad})E\sub{int}=\frac{3\gamma\sub{ad}-4}{3(\gamma\sub{ad}-1)}E\sub{g}\,.
\label{eq:Etot}
\end{equation}
%%%
If we assume the polytropic relation that $P(\rho)=K\rho^{1+1/n}$, 
%%%
\begin{equation}
E\sub{g}=-\frac{3}{5-n}\frac{GM_\star^2}{R_\star}~,
\end{equation}
%%%
where $n$ is the polytropic index. 
If we define $C\equiv(3\gamma\sub{ad}-4)/(\gamma\sub{ad}-1)(5-n)$, $E\sub{tot}=-CGM_\star^2/R_\star$. 
Thus, the total energy evolution in Eq.~(\ref{eq:dEdt}) is transformed as
%%%
\begin{equation}
\frac{\dot{R}}{R_\star} = \left( 2-\frac{1-\xi}{C}\right) \frac{\dot{M}}{M_\star} -\frac{RL_\star}{CGM^2}+\frac{RL\sub{nuc}}{CGM^2}~.\label{eq:KHMacc}
\end{equation}
%%%

The second term of RHS of Eq.~(\ref{eq:KHMacc}) corresponds to the inverse of the {{K-H}} timescale $\tau\sub{KH}$, which is defined as the typical timescale of the radiative cooling (i.e., $\tau\sub{KH} \equiv|E\sub{tot}|/L_\star$). 
During the main-accretion phase{{, $\tau\sub{KH}$}} is in general much longer than the accretion timescale $\tau\sub{acc} \equiv M/\dot{M}$ \footnote{This is equivalent to the condition $L_\star \ll L\sub{acc}$ as shown in Fig.~\ref{fig:1d-5-t-L}.}.
Therefore, the second term of RHS of Eq.~(\ref{eq:KHMacc}) can be neglected if the first term of RHS is not zero, i.e., $1-\xi \neq 2C$.

In fully convective stars, which consist of the monatomic ideal gas, $n=3/2$ and $\gamma\sub{ad}=5/3$ and then $C=3/7$. In addition, in the case that $\xi=0$, we obtain Eq.~(\ref{eq:lnuccrit}).
Moreover if $L\sub{nuc}\ll L\sub{acc}$, it is transferred as Eq.~(\ref{eq:R-M}).

\section{Dependence on the initial stellar seed mass} \label{app:Mini}

In this paper we chose $10^{-2}~\Msun$ as the fiducial value of the initial stellar seed mass.
We stress that this value is {{higher than the second-core mass in  recent work of the hydrodynamic collapse of molecular clouds 
\citep[$\sim \Mjup$; see][]{Masunaga+Inutsuka00,Stamatellos+07,Tomida+13,Vaytet+13}.}}
For example, a second core is formed with $4\,\Mjup$ and $4\,\Rsun$ in \citet{Masunaga+Inutsuka00}, while $1.4\,\Mjup$ and $0.65\,\Rsun$ in the ten simulations of \citet{Vaytet+13} \footnote{\citet{Vaytet+13} indicate that the difference probably results from different opacity tables.}.
Therefore, our calculations start with slightly evolved seeds rather than second cores.

Our approach in {{Sects.~\ref{sec:cold-ini} and \ref{sec:ini}}} was to fix the initial seed mass and explore different values of the initial seed radius, following \citetalias{Hosokawa+11}. This choice was essentially motivated by the fact that the convergence of evolution models with MESA is more difficult for very low seed masses. 
However, in light of the fact that, as pointed out by \citetalias{BVC12}, the low value of the initial seed mass has important consequences for the subsequent stellar evolution, we must show that the range of initial conditions yield evolutionary tracks that are equivalent to those that we would obtain with small initial seed masses.

\begin{figure}[!tb]
  \begin{center}
    \includegraphics[width=\hsize,keepaspectratio,clip]{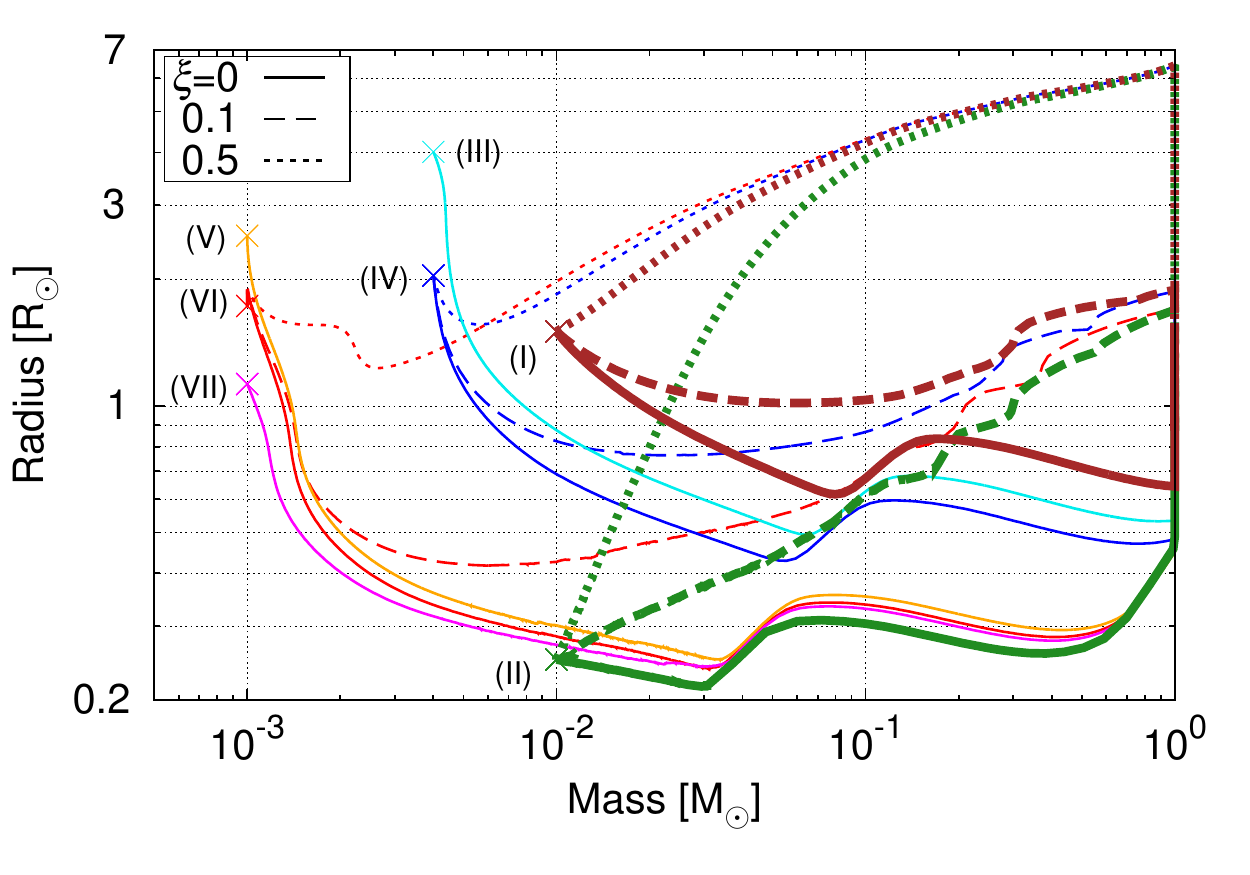}
        \caption{\small{
        Evolution of radius as a function of mass for accreting stars with $\dot{M}=10^{-5}~\Msun/\rm{yr}$ for different initial conditions and different values of $\xi$. 
        The crosses (with corresponding I to VII labels) correspond to the initial conditions chosen to span a range of specific entropies. The different initial radii are $1.5\,\Rsun$ (case I), $0.25\,\Rsun$ (II), $4.0\,\Rsun$ (III), $2.0\,\Rsun$ (IV), $2.5\,\Rsun$ (V), $1.8\,\Rsun$ (VI), and $1.1\,\Rsun$ (VII).
        Solid, dashed, and dotted lines indicate the evolution with $\xi=0, 0.1$ and $0.5$, respectively. The evolution for cases (I) and (II) correspond to those adopted in the present manuscript and are highlighted with thicker lines. 
                }}\label{fig:Mini-R}
    \end{center}
\end{figure}

Figure~\ref{fig:Mini-R} illustrates seven evolutionary models starting from different seed masses: $0.01~\Msun$ (I--II),  $0.004~\Msun$ (III--IV),  and $10^{-3}~\Msun$ (V--VII), for various initial radii $R_{\rm ini}$ and heat injection efficiencies $\xi$. 
Case (I) corresponds to the fiducial initial condition used in the present paper and case (II) is used in Sect.~\ref{sec:ini}.
Case (III) corresponds to seed conditions obtained by \citet{Masunaga+Inutsuka00}. Case (IV) corresponds to a seed with the same specific entropy as our fiducial seed (I) but a mass of only $0.004~\Msun$ (see Eq.~\ref{eq:R-M}).
Case (V) corresponds to the largest specific entropy for which we could converge an evolution calculation for a $10^{-3}~\Msun$ seed. 
The radius of a $10^{-3}~\Msun$ seed with the same entropy as case (I) would be $3.2~\Rsun$, but unfortunately the corresponding calculation failed to converge. Finally, cases (VI) and (VII) are obtained by arbitrarily changing the initial radii for $10^{-3}~\Msun$ seeds to span a range of specific entropies. 
{{Interestingly, for stars evolving from the initial conditions (II), (V), (VI), or (VII), their radius is so small that they reach high enough central temperatures to ignite hydrogen burning when their mass reaches $\simeq0.6\,\Msun$.}}

%%%%%%
%%%%%%

We can see in  Fig.~\ref{fig:Mini-R} that the range of radii obtained for models with different initial seed masses and different values of $\xi$ is the same as that obtained with the our fiducial seed mass and various values of the initial radius. Furthermore, we see that the evolution is largely controlled by the value of the $\xi$ parameter: For $\xi=0.5$, the evolutionary tracks converge independent of the choice of the initial seed properties above about $0.1\,\Msun$. This is also the case for the $\xi=0.1$ case, although differences in radii on the order of $\sim 10\%$ remain even after accretion is completed. The cold accretion case is the one for which differences in initial conditions persist the longest and can still be on the order of $\sim 40\%$ at the end of the accretion phase.  Our conclusion that most stars would have been formed with $\xi\gtrsim0.1$ is not affected by the uncertainties on the initial conditions.

In order to explain the small sizes of young cool stars, \citetalias{Hosokawa+11} invoked the need for small seed radius ($0.2~\Rsun$ at $0.01~\Msun$) while \citetalias{BVC12} invoked the need for a small seed mass. 
Our calculations in Fig.~\ref{fig:Mini-R} show that these conditions are equivalent.

\end{appendix}

%\Online

\end{document}